\documentclass[fleqn,usenatbib]{mnras}

\usepackage{newtxtext,newtxmath}

\usepackage[T1]{fontenc}

\DeclareRobustCommand{\VAN}[3]{#2}
\let\VANthebibliography\thebibliography
\def\thebibliography{\DeclareRobustCommand{\VAN}[3]{##3}\VANthebibliography}

\usepackage{soul}

\usepackage{graphicx}	
\usepackage{amsmath}	
\usepackage{subcaption}
\usepackage{multirow}

\title[Links Between Optical and X-ray Light in Cygnus X-2]{Links Between Optical and X-ray Light in Cygnus X-2}

\author[A. B. Igl et al.]{
Alexander B. Igl,$^{1}$
R. I. Hynes,$^{1}$\thanks{E-mail: rhynes@lsu.edu}
K. S. O'Brien,$^{2}$
E. L. Robinson$^{3}$
C. T. Britt,$^{4}$
\\
$^{1}$Department of Physics and Astronomy, Louisiana State University, 202 Nicholson Hall, Tower Drive, Baton Rouge, Louisiana 70803, USA\\
$^{2}$Department of Physics, Durham University, Odgen Centre For Fundamental Physics West, Lower Mountjoy, South Rd, Durham DH1 3LE, United Kingdom\\
$^{3}$Department of Astronomy, University of Texas at Austin, 1 University Station, C1400, Austin, Texas 78712\\
$^{4}$Space Telescope Science Institute, 3700 San Martin Drive, Baltimore, Maryland 21218\\
}

\date{Accepted XXX. Received YYY; in original form ZZZ}

\pubyear{2023}

\begin{document}
\label{firstpage}
\pagerange{\pageref{firstpage}--\pageref{lastpage}}
\maketitle

\begin{abstract}
We observed the low mass X-ray binary Cyg X-2 for a total of 18 nights over two observing runs in July and September of 2006, using the Otto Struve Telescope at McDonald Observatory and the Rossi X-ray Timing Explorer.  Using discrete cross correlations, we found peaks occurring at near-zero lags in the flaring branch of the colour-colour diagram, which could signify reprocessing, in addition to an anti-correlation within the normal branch.  When comparing optical flux to the system's placement on the Z track, two distinct behaviors were seen: (1) a state with no correlation, and (2) a multi-valued (horizontal and normal branches)/correlated (flaring branch) state.  The correlation was the result of direct steps and more gradual falls to and from the flaring branch respectively.  Finally, we modeled timed spectra with 64 second bins with an extended accretion disc corona model.  We found that correlations occurred between the optical and the various fitted parameters, particularly the blackbody normalization (and blackbody radius by extension) in higher intensity regions.  Despite this, the Z track location was found to be a far better predictor of physical parameters than the optical flux, with clean correlations seen in every branch of the Z track.  Where optical correlations are found, the location on the Z track was a better predictor of optical flux than any individual physical parameter.
\end{abstract}




\section{Introduction}

An X-ray binary (XRB) system is a semi-detached binary system that contains a normal star or white dwarf donor and a neutron star (NS) or black hole (BH) accretor.  Through processes such as Roche lobe overflow, winds, or circumstellar disc interaction, material streams down into an accretion disc surrounding the central object.  At the inner radii of the accretion disc, there is enough viscous heating for the material to emit X-rays.  Depending on the mass of the donor, an XRB can be divided into low mass ($<1M_ \odot$) or high mass ($>10M_ \odot$) classes.

Neutron star low mass XRBs (LMXBs) can be divided into two subclasses, known as Z types and atolls \citep{1989A&A...225...79H}, depending on the shape of their X-ray colour-colour diagrams (CDs) and hardness-intensity diagrams (HIDs).  Atolls have three main branches: the island state takes the shape of a formless clump, and the lower and upper banana branches together create a curve shaped like their namesake.  Z sources are shaped like the letter, and have three branches as well: the horizontal branch (HB), the normal branch (NB), and the flaring branch (FB).  They tend to be brighter than their atoll counterparts, generally being near or above the Eddington limit, but also less numerous, with only six persistent Z sources having been confirmed in our Galaxy.  Z sources also move through all three states more quickly than atolls, on the orders of hours to days and weeks respectively \citep{1989A&A...225...79H}.  Finally, both see their shapes change and translate on the CDs, something known as secular drift.  Cyg X-2 in particular has been observed to have a high amount of secular drift \citep{1996A&A...311..197K}.

The main driver of where LMXBs lie on their CDs is generally thought to be the mass accretion rate.  Traditionally, it has been thought to increase in Z sources from HB to FB, but there are a number of proposed alternatives.  \citet{2006A&A...460..233C} suggests that minimum accretion occurs between the NB and FB (the soft apex), while \citet{2009ApJ...696.1257L} claims that other physical properties drive the location, and that mass accretion rate is constant across the whole track.

Z sources can be further broken down into two subcategories \citep{1994A&A...289..795K}: Sco-likes and Cyg-likes.  Sco-likes (Sco X-1, GX 17+2, GX 349+2) are named after Sco X-1, and have a Z track shaped more like a Greek ``$\nu$'', with a short, upturned HB, and a prominent FB.  Cyg-likes (Cyg X-2, GX 340+0, GX 5-1) are named after Cyg X-2, and have a Z track that looks more like the letter ``Z'', with a more prominent HB and smaller FB than the Sco-likes.  A number of mechanisms have been proposed to explain the differences between the aforementioned subtypes, as well as (more generally) Z types and atolls.  Suggestions included inclination angles, NS spin rates, NS masses, and NS magnetic field strengths \citep{1989A&A...225...79H}.  The discovery of XTE J1701-462 \citep{2006ATel..696....1R} put these options to rest, and suggests that accretion rate is what determines XRB type.  Because it went through all three source states over its 19 month outburst \citep{2007ApJ...656..420H,2007ATel.1144....1H}, the necessary timescales for changes of the previously mentioned physical parameters disqualifies them.  Along with J1701-462, other LMXBs have been seen to transition between multiple states.  GX 13+1 was seen by \citet{2015ApJ...809...52F} to transition between Cyg-like and Sco-like states, and \citet{2022MNRAS.513.2708R} found that Swift J1858.6-0814 bridged the atoll/Z source gap on the radio/X-ray plane.

It is well established that the spectra of neutron star LMXBs have both a thermal component and a non-thermal component.  However, there is not yet a single agreed upon model.  Two models in particular are often cited: the Eastern model \citep{1989PASJ...41...97M} and the Extended Accretion Disc Coronal (ADC) model \citep{1995A&A...300..441C,2004MNRAS.348..955C}.  In the Eastern model, soft X-ray emission would be dominated by a multicoloured blackbody disc. The hard component comes from the Comptonization of seed photons from the inner disc in a spherical corona around the NS. On the other side of the coin, the extended ADC model also has soft emission dominated by a blackbody, this time originating from on or near the NS surface. Comptonization occurs in a corona that exists as a layer above the inner radii of the accretion disc. An additional line component is often necessary for a good fit. A contribution from the K$\alpha$ transition in iron is usually found in the mid-6 keV range.

Dipping sources, which are LMXBs viewed at an angle between 65--85$^\circ$, may be the key to discerning which model is correct.  The region on the disc that touches the accretion stream is puffed up, creating a thick, absorbing bulge at the rim which extends up to 70$^{\circ}$ in azimuth \citep{1982ApJ...253L..61W}.  This leads to orbital-related brightness dips as the bulge covers an observer's line of sight to the corona.  Because this is a purely geometric effect, models are more strongly constrained, as they should fit well both inside and outside of the dips.  Dipping sources have already been used to get a clearer picture of the geometry of an extended ADC.  \citet{2004MNRAS.348..955C} showed that the ADC would extend from 20,000 km for faint sources to 700,000 km for bright ones, since the Comptonized component is obscured very slowly with dipping.  The fractions of extended ADC radius to disc radius ranged from 6.4\% to 64.8\%, so the ADC has been seen to cover a significant portion of the accretion disc.  The ADC also seems to be geometrically thin ($H/r << 1$), because 100\% deep dipping would not be seen otherwise.  Cyg X-2 itself has shown evidence of an extended ADC.  \citet{1988ApJ...329..276V} found that variations in Cyg X-2 dips line up with the scenario where the dips are dependent on the thickness of the disc and ADC (both geometric and optical).  Another example is given in \citet{2009ApJ...692L..80S}, who used line widths and the ratios of forbidden and intercombination line fluxes to determine a lower limit of (6.4$\pm$1.4)$\times$10$^{14}$cm$^{-3}$ for the density of the corona.

The namesake of the Z source subclass it belongs to, Cygnus X-2 was first discovered in the X-ray by \citet{1966Sci...152...66B} using a sounding rocket.  It is comprised of a neutron star and a late type (A9) companion, circling each other in an $\sim$9.8 day orbit \citep{1979ApJ...231..539C}.  \citet{1999MNRAS.305..132O} found an inclination of $62.5\pm4^\circ$, and NS and companion masses of $1.78\pm0.23M_\odot$ and $0.60\pm0.13M_\odot$ respectively. \citet{1979ApJ...231..539C} found the source to be at a distance of $\sim$8 kpc, but more recently, \citet{2021PASA...38...48D} used Gaia EDR3 data to calculate a distance of $11.3^{+0.9}_{-0.8}$ kpc.

Because Cyg X-2 is so luminous, the system has been well studied spectrally.  It is well established that its spectra contain an iron K$\alpha$ line at $\sim$6.7 keV \citep{1993ApJ...410..796S}, which likely results from reflection on the disc.  In addition, the system is known to have spectral features in the 0.2-1.5 keV range that are consistent with plasma that is collisionally excited, or possibly photoionized \citep{1986ApJ...307..698V}.  Spectral modeling of NuSTAR data using a relativistically blurred reflection model in \citet{2018MNRAS.474.2064M} found large inner disc radii of $\sim$24-32 km in the dipping spectrum and a radius of $\sim$30-73 km in the non-dipping.

In the HID of Cyg X-2, the intensity usually decreases on the FB, creating a backwards ``C'' (Cyg-like) shape \citep{2022ApJ...927..112L,2004MNRAS.350..587O}.  However, intensity has also been seen to increase on the FB, making a ``Z'' shaped HID, as in \citet{2010A&A...512A...9B}.  The authors of this paper say that the difference lies in the distinction between dipping and flaring, where the true FB is an intensity increase, and occurs rarely \citep{2012MmSAI..83..178B}.  \citet{2015ApJ...809...52F} disagree with the interpretation that the dipping and flaring branches are caused by different mechanisms (nuclear burning and outer disc absorption), as they noticed the change between the two occurring as a smooth transition (rotation) from one to the other.  Instead, the authors suggest that the accretion rate is the primary driver of the dipping/flaring branch differences.  They note that when the Z track lies at higher intensities, the shape becomes Cyg-like, and when the Z track lies at lower intensities, the shape becomes Sco-like.

In binary systems such as XRBs, X-ray photons from the inner disc radii can photoionize material on the outer disc or companion star.  When the electrons recombine, they release a photon of lower energy than the initial X-ray (optical/UV), a process known as thermal reprocessing.  Reprocessing is thought to account for much of the optical light seen in LMXBs \citep{1994A&A...290..133V}.  Over time, an entire X-ray signal can be reprocessed into the optical regime, where the optical lightcurve is observed with a delay on the order of the light travel time through the system ($\sim$1-10 s for most LMXBs).  The timescale of reprocessing itself is nonzero, although it is negligible compared to light travel times \citep{1987ApJ...315..162C}.

The reprocessed signal is given by
\begin{equation}
    F_{\rm o}(t)=\int \Psi (\tau) F_{\rm x}(t- \tau)d \tau
\end{equation}
where $F_o$ is the optical flux, $F_x$ is the X-ray flux, $t$ is the time, and $\tau$ is the optical time delay, and $\Psi$ is the transfer function, which describes the amount of X-ray light that is reprocessed into the optical as a function of lag.  The possible lag times are based on the geometry of the system, ranging from a minimum of zero and a maximum of $2R/c$, where $R$ is the furthest reprocessing site on the system, and $c$ is the speed of light.  Lags in between can be shown in the form of isodelay surfaces, summarized in \citet{2003SPIE.4854..262H}.  \citet{2002MNRAS.334..426O} modeled the transfer function of a typical LMXB as a function of lag and phase, which revealed two main components.  The first is from the disc, which occurred at lower lags, and was constant in phase.  The second was from the companion star, which was quasisinusoidal and at higher lags.  The accretion stream would also be included in the second component, but its contribution is relatively small due to the small area.

In this paper, we analyze how the X-ray regime is related to the optical from a number of angles.  In Section \ref{sect:obs}, we elaborate on the conditions used for data acquisition.  Section \ref{sect:data} contains descriptions of trends and features seen in the lightcurves and CDs.  Section \ref{sect:z_param} discusses the process for Z track parameterization.  Section \ref{sect:ccf} describes the search for thermal reprocessing using cross correlation functions, and the results of that search.  Section \ref{sect:z_behave} shows the results of the Z track parameterization.  Section \ref{sect:spectral} contains the results for the timed spectral fits that were performed on the X-ray data, and how they are related to the optical and the Z track location.  Finally, Sections \ref{sect:discussion} and \ref{sect:conclusions} are the discussion of results and conclusions respectively.

\section{Observations}
\label{sect:obs}
\begin{table}
    \centering
    \begin{tabular}{ll} 
        \hline
        Optical Range (UT) & X-ray Range (UT) \\
        \hline
        July 25.68 - 25.98, 2006 & July 25.68 - 25.94, 2006 \\
        \hline
        July 26.65 - 26.99, 2006 & July 26.66 - 26.94, 2006 \\
        \hline
        July 27.65 - 27.98, 2006 & July 27.64 - 27.92, 2006 \\
        \hline
        July 28.67 - 28.85, 2006 & July 28.62 - 28.94, 2006 \\
        \hline
        July 29.66 - 29.98, 2006 & July 29.67 - 29.94, 2006 \\
        \hline
        July 30.65 - 30.70, 2006 & July 30.65 - 30.93, 2006 \\
        \hline
        N/A & July 31.63 - 31.91, 2006 \\
        \hline
        August 1.68 - 1.76, 2006 & August 1.61 - 1.89, 2006 \\
        \hline
        September 23.58 - 23.88, 2006 & September 23.71 - 23.81, 2006 \\
        \hline
        September 24.58 - 24.90, 2006 & September 24.75 - 24.86, 2006 \\
        \hline
        September 25.56 - 25.87, 2006 & September 25.74 - 25.84, 2006 \\
        \hline
        September 26.56 - 26.91, 2006 & September 26.72 - 26.82, 2006 \\
        \hline
        September 27.58 - 27.92, 2006 & September 27.70 - 27.80, 2006 \\
        \hline
        September 28.58 - 28.91, 2006 & September 28.75 - 28.85, 2006 \\
        \hline
        September 29.69 - 29.91, 2006 & September 29.73 - 29.83, 2006 \\
        \hline
        September 30.57 - 30.89, 2006 & September 30.65 - 30.74, 2006 \\
        \hline
        October 1.65 - 1.75, 2006 & October 1.69 - 1.79, 2006 \\
        \hline
        October 2.59 - 2.84, 2006 & October 2.81 - 2.84, 2006 \\
        \hline
    \end{tabular}
    \caption{Time ranges of all data.}
    \label{table:data_tr}
\end{table}

The optical observations were performed over two runs at McDonald Observatory, using the Argos instrument on the Otto Struve Telescope \citep{2005JApA...26..321M}.  The first was from July 25, 2006 to August 1, 2006, and the second from September 23, 2006 to October 2, 2006.  Data were taken every night, except on July 31 due to weather.  The timeranges with data can be viewed in Table \ref{table:data_tr}.  A broad BVR filter was used to make lightcurves with 1 s time resolution.  Using the bias, dark, and flat frames taken nightly, custom IDL codes written for Argos were used to reduce the data.  Most of the lightcurves were $\sim$7 hours of continuous data, although due to weather, seeing, etc., some of the lightcurves are shorter or are taken with some data gap, like in Fig. \ref{fig:54002_lc}.  Syncing the absolute time was done using Network Time Protocol (NTP) servers and Global Positioning System (GPS) 1 s ticks.  To reduce the the effects of the atmosphere, all lightcurves used were differential, using 2MASS J21444211+3817558 as the comparison star.  If atmospheric effects became too strong, even differential lightcurves were not enough to mask them.  As such, a minimum count threshold was set on an observation by observation basis, so that these bad data could be ignored.

The X-ray data were taken with the Rossi X-ray Timing Explorer (RXTE) Proportional Counter Array (PCA).  The PCA was comprised of five Proportional Counter Units (PCUs), with a total collecting area of 6500 cm$^2$ and a usable energy range of 2-60 keV.  Some of the PCUs began to discharge with instrument aging.  In an effort to prevent further damage, some PCUs are turned off for periods of time.  This leads to noticeable changes in count rates, which needs to be accounted for, specifically in the STANDARD-1 data. The STANDARD-2 data were reduced using only the count rate from PCU2 to ensure consistency.  We used both STANDARD-1 and STANDARD-2 data for this study.  The STANDARD-1 data, which do not contain energy bands, were taken with 1 s time resolution, matching the Argos data.  The STANDARD-2 data were taken with 16 s resolution.  We extracted lightcurves in the energy bands 2.06-3.68 keV, 3.68-6.12 keV, 6.12-8.98 keV, and 8.98-14.76 keV.  Soft colour is defined as the ratio of the 3.68-6.12 keV band over the 2.06-3.68 keV band, and hard colour is defined as the 8.98-14.76 keV band over the 6.12-8.98 keV band.

The spectra were reduced from the raw outputs using standard tools in HEASoft (version 6.29), based on the instructions of the ``New PCA Tools: Overview'' recipe in the RXTE Cookbook\footnote{https://heasarc.gsfc.nasa.gov/docs/xte/recipes2/Overview.html}.  First, the raw data are prepared for analysis through processes such as creating filter files and estimating the background.  Next, the good time filter is set, which uses user set conditions to determine what data to keep.  The conditions used in the extraction were (1) ensuring that the target was above the Earth's horizon (ELV $>$ 4), (2) making sure the PCA is pointing within 0.1 degrees of the target (OFFSET $<$ 0.1), and (3) keeping only times where at least one PCU was active (NUM$\textunderscore$PCU$\textunderscore$ON $>$ 0), but getting rid of unreal active PCU numbers (NUM$\textunderscore$PCU$\textunderscore$ON $<$ 6).  All of these are considered basic screening recommendations.  The X-ray spectral data were extracted for 64 s bins.  These bins were aligned with the times in the STANDARD-2 lightcurves, so that they could be matched with the X-ray observables and binned optical data to create timed spectra.

\section{Data Description}
\label{sect:data}
\begin{figure}
	\includegraphics[width=\columnwidth]{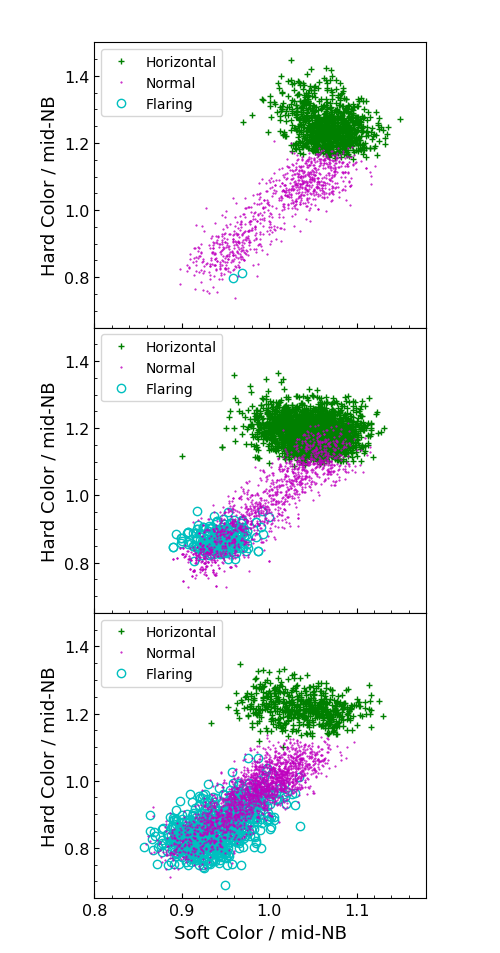}
    \caption{The CDs for each time separated group, where the axes are normalized to the NB midpoint ($s_z$=1.5) for each group.  Each group corresponds to a different date range (see Table \ref{table:groups}).}
    \label{fig:ccd}
\end{figure}

\begin{figure}
	\includegraphics[width=\columnwidth]{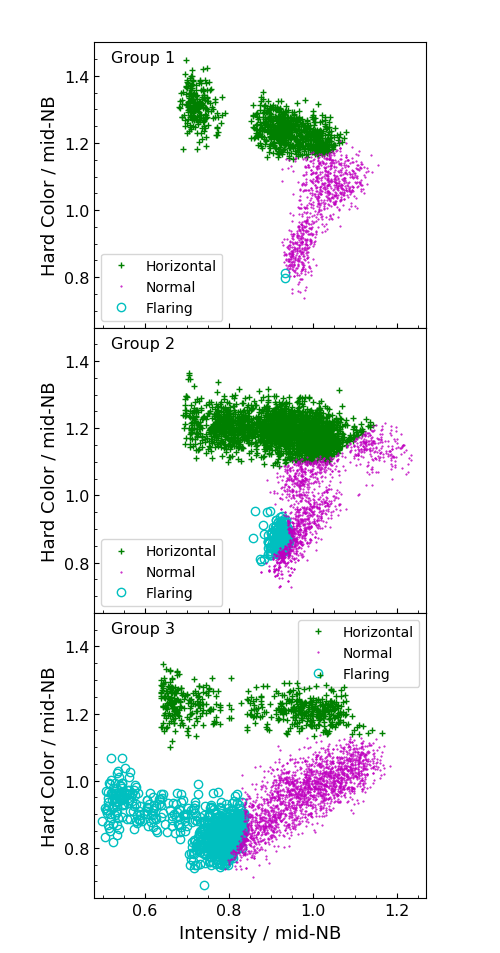}
    \caption{The HIDs for each time separated group, where the axes are normalized to the NB midpoint ($s_z$=1.5) for each group.  Each group corresponds to a different date range (see Table \ref{table:groups}).}
    \label{fig:hid}
\end{figure}

The CD has some significant secular drift, as may be expected for Cyg X-2.  As such, it is separated into three groups (summarized in Table \ref{table:groups}, CDs in Fig. \ref{fig:ccd}, HIDs in Fig. \ref{fig:hid}), two of which are in July and the other comprising of all September data.  These groups will from now on be referred to as groups 1, 2, and 3, in order of time the data were taken (see Table \ref{table:adjust} for the parameters needed to align the groups).  Within each group, drift much is lower, with a maximum of about 1000 counts/s on the intensity axis.  The amount of data differs by group, as does the Z track coverage, where groups 2 and 3 each have almost twice as much data as group 1.  Groups 1 and 2 have nearly identical branch coverage, stretching through the HB to the soft apex.  Group 3 has less HB data, but it is the only group that includes a significant amount of data in the FB.

\begin{table}
    \centering
    \begin{tabular}{lll} 
        \hline
        Group Number & Time Range (UT) & MJD (UT) \\
        \hline
        1 & July 25 - July 27, 2006 & 53941 - 53943 \\ 
        \hline
        2 & July 28 - August 1, 2006 & 53944 - 53948 \\
        \hline
        3 & September 23 - October 2, 2006 & 54001 - 54010 \\
        \hline
    \end{tabular}
    \caption{Time ranges of the group separations used to compensate for secular drift.}
    \label{table:groups}
\end{table}

\begin{figure}
    \setlength\tabcolsep{0pt}
    \begin{tabular}{c}
        \begin{subfigure}[b]{.475\textwidth}
            \setcounter{subfigure}{0}
            \includegraphics[width=\textwidth]{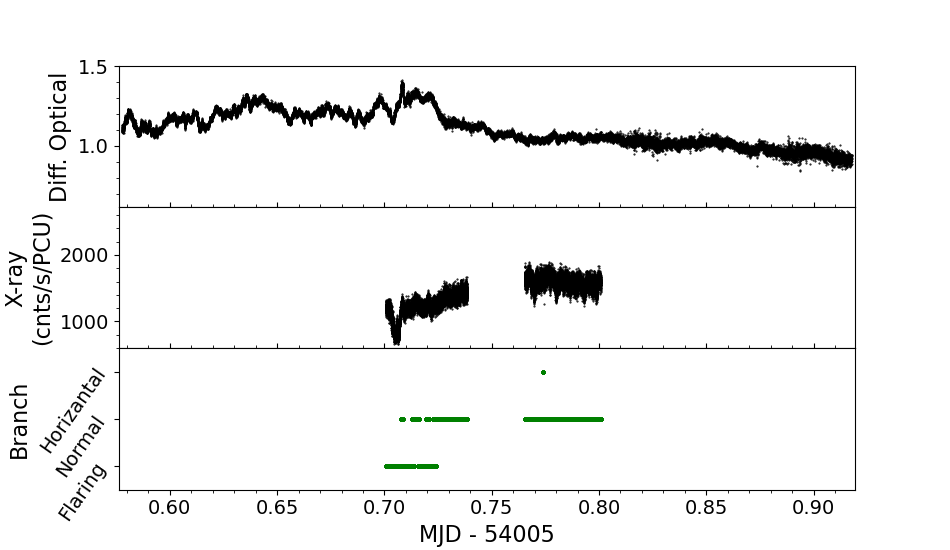}
            \caption[empty]{}
            \label{fig:54005_lc}
        \end{subfigure}\\
        \hfill
        \begin{subfigure}[b]{.475\textwidth}
            \setcounter{subfigure}{1}
            \includegraphics[width=\textwidth]{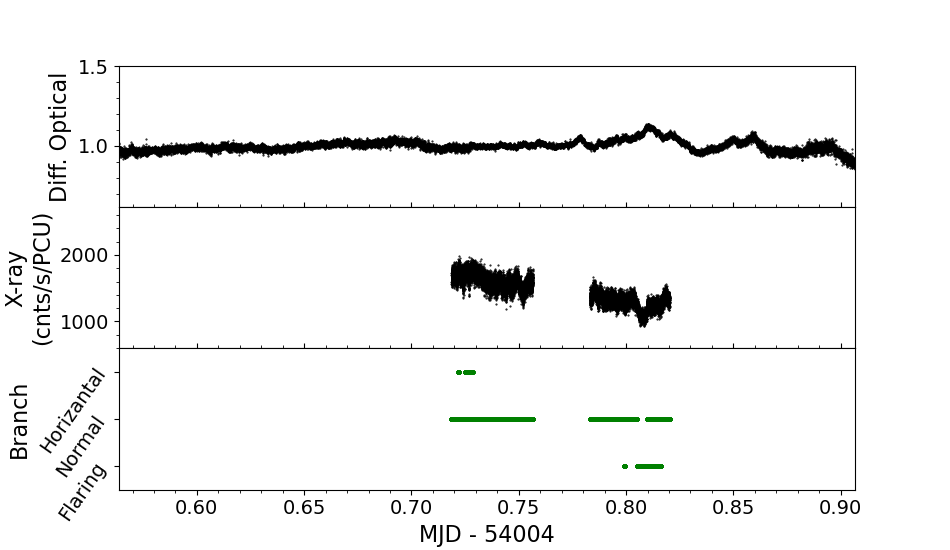}
            \caption[empty]{}
            \label{fig:54004_lc}
        \end{subfigure}\\
        \hfill
        \begin{subfigure}[b]{.475\textwidth}
            \setcounter{subfigure}{2}
            \includegraphics[width=\textwidth]{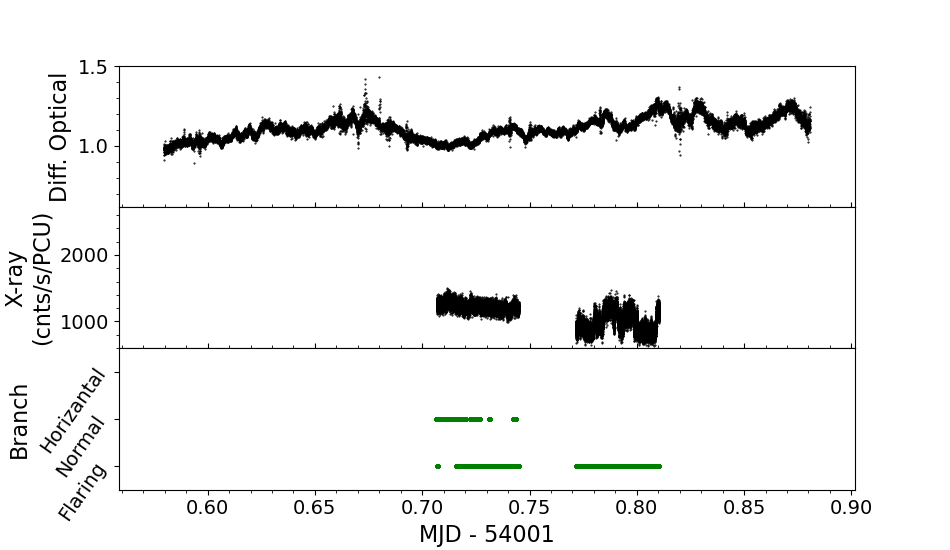}
            \caption[empty]{}
            \label{fig:54001_lc}
        \end{subfigure}
    \end{tabular}
    \caption{In each subplot, the top panel is the differential optical lightcurve, the middle panel is the optical lightcurve, and the bottom panel is the Z track location. Fig. \ref{fig:54005_lc} is the data taken on MJD 54005.  The X-ray data contains examples of hour-long features and increased variability while the system is on the FB.  Fig. \ref{fig:54004_lc} is the data taken on MJD 54004.  The X-ray lightcurve contains an example of increased variability while the system is on the FB.  The effect described in Section \ref{sect:z_behave} has adversely impacted the branch location plot here, as $s_z$ ranges from 1.25 - 2.15.  Fig. \ref{fig:54001_lc} is the data taken on MJD 54001.  The system is most active here, alternating between count rates of $\sim$750 and 1100 counts/s/PCU.}
\end{figure}

\begin{figure}
	\includegraphics[width=\columnwidth]{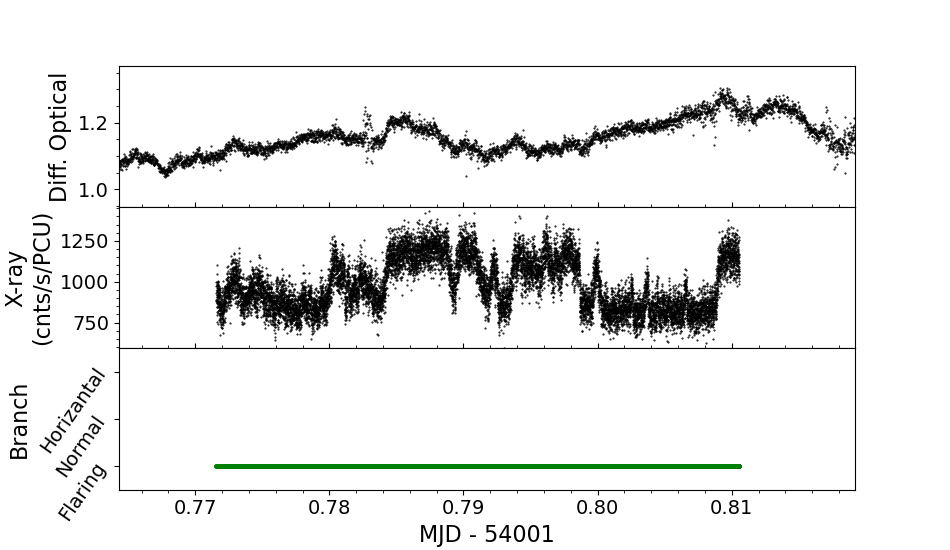}
    \caption{A zoomed version of the MJD 54001 data (Fig. \ref{fig:54001_lc}) for identification of X-ray lightcurve features.}
    \label{fig:54001_zoom}
\end{figure}

\begin{figure}
	\includegraphics[width=\columnwidth]{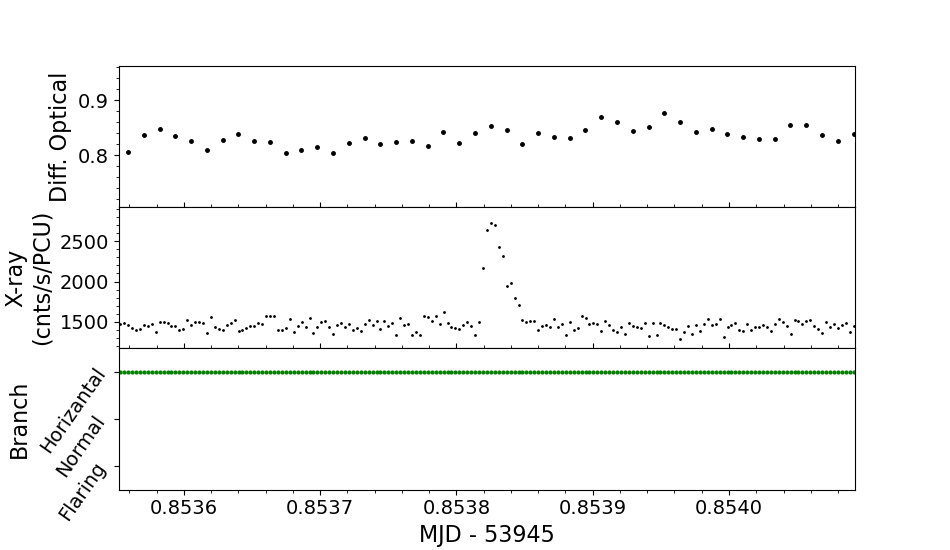}
    \caption{The X-ray burst seen on MJD 53945.  This is a subset of the lightcurve seen in Fig. \ref{fig:53945_lc}.}
    \label{fig:xr_burst}
\end{figure}

The X-ray data comprise segments that generally last about an hour (one per Earth orbit of the satellite, see Appendix \ref{lc_apx}).  As is seen in the HIDs, most of the data are in the HB or NB, and very little FB exists in the dataset.  This could contribute to the lack of large scale changes seen within the data.  However, some hour-long timescale changes can be seen, like in Fig. \ref{fig:54005_lc}, where the count rate nearly doubles over the course of a couple of hours.  Where Cyg X-2 does venture into the FB, the lightcurve becomes more variable.  For example, at MJDs $\sim$54004.81 (Fig. \ref{fig:54004_lc}) and $\sim$54005.70 (Fig. \ref{fig:54005_lc}), the count rate briefly decreases and returns to near its original value in $\sim$7 minutes.  The system is the most active on MJD 54001 (Fig. \ref{fig:54001_lc}), where the lightcurve alternates between count rates of about 750 and 1100 counts/s/PCU.  Interspersed among these steps are the previously mentioned quick dips (MJD 54001.789, Fig. \ref{fig:54001_zoom}), as well as spikes of similar duration (MJD 54001.792, 54001.800, Fig. \ref{fig:54001_zoom}).  The dips appear when the count rate is high, and the spikes when it is low.  Although the transition between the steps is usually quick, one slower decrease can be seen starting at MJD 54001.780 (Fig. \ref{fig:54001_zoom}).  It is worth mentioning that this behavior looks very similar to the behavior in the Kepler K2 lightcurves of Sco X-1 in \citet{2016MNRAS.459.3596H}, which was noted to be bimodal.  The dips and spikes in Fig. \ref{fig:54001_zoom} could then be thought of as quick reversions to and from the opposite state.  One very clear X-ray burst is seen on MJD 53945 (Fig. \ref{fig:xr_burst}, see Fig. \ref{fig:53945_lc} for full lightcurve), taking place on the HB.

\begin{table}
    \centering
    \begin{tabular}{llll} 
        \hline
        \multirow{6}{*}{Group 1} & \multirow{2}{*}{Soft} & Shift & 0.03 \\ 
                                 &                       & Scale & 0.7 \\
                                 & \multirow{2}{*}{Hard} & Shift & 0.12 \\
                                 &                       & Scale & 1.0 \\
                                 & \multirow{2}{*}{Flux} & Shift & 0.14 \\ 
                                 &                       & Scale & 1.0 \\
        \multirow{6}{*}{Group 2} & \multirow{2}{*}{Soft} & Shift & 0.0 \\ 
                                 &                       & Scale & 1.0 \\
                                 & \multirow{2}{*}{Hard} & Shift & 0.0 \\
                                 &                       & Scale & 1.5 \\
                                 & \multirow{2}{*}{Flux} & Shift & 0.0 \\ 
                                 &                       & Scale & 1.0 \\
        \multirow{6}{*}{Group 3} & \multirow{2}{*}{Soft} & Shift & 0.03 \\ 
                                 &                       & Scale & 0.8 \\
                                 & \multirow{2}{*}{Hard} & Shift & 0.09 \\
                                 &                       & Scale & 1.3 \\
                                 & \multirow{2}{*}{Flux} & Shift & 0.21 \\ 
                                 &                       & Scale & 0.3 \\
        \hline
    \end{tabular}
    \caption{Adjustments made to group data for ($s_z$,$t_z$,$u_z$) coordinate mapping.  The shifts have been normalized to the NB midpoint, for comparison with Figs. \ref{fig:ccd} and \ref{fig:hid}.}
    \label{table:adjust}
\end{table}

\section{Z Track Parameterization}
\label{sect:z_param}
To quantify the state evolution, the Z track needed to be parameterized.  The scheme developed is very similar to the one described in \citet{2000MNRAS.311..201D}, and is further discussed in Appendix A of \citet{igl_scox1}.  The main difference is that instead of visualizing the Z track in 2D, it uses a 3D colour-colour-intensity (CCI) diagram.  Thus, the parameters ($s_z$,$t_z$,$u_z$) are created, where $s_z$ represents the location on the Z track, $t_z$ represents the distance away from the Z track, and $u_z$ represents the angular component around the track.  An accurate analogy would be to think of a ``warped'' cylindrical coordinate system, where ($s_z$,$t_z$,$u_z$) are similar to ($z$,$r$,$\phi$) respectively.  Using the CCI for the ranking is especially important for Cyg X-2, as the FB doubles back onto the NB in colour-colour space alone, but has a distinct intensity behavior.

One roadblock to a good mapping was the significant secular drift that occurs in the CD.  Of course, the parameterization could be performed separately on each group, but this would make direct comparison between the behaviors of the three much more difficult.  To get around this, the CDs of each group were transformed through scaling and shifting, until all three appeared to be part of the same Z track (parameters in Table \ref{table:adjust}).  The mapping could then be transformed back, and used with the original data.  Another issue that arises is the scale of each CCI axis.  The intensity scale ($\sim10^3$ counts/s) is much larger than the colour scales ($\sim10^{-1}-10^0$), and so left as is, the intensity would dominate the mapping.  This problem was avoided by normalizing the axes to the NB-midpoint before the mapping.

\begin{figure}
	\includegraphics[width=\columnwidth]{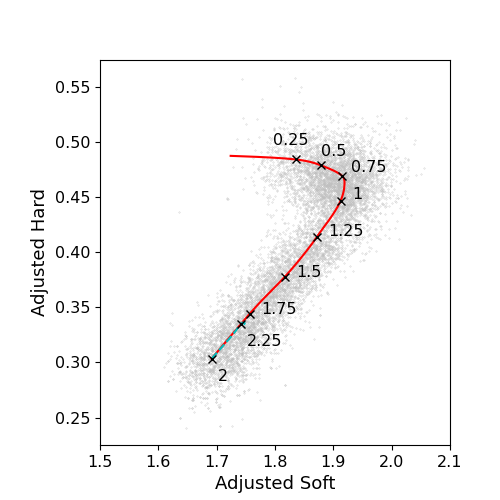}
    \caption{Selected $s_z$ values are labeled on the interpolated Z track, along with the transformed data used to create them.  The dashed cyan line represents the FB.}
    \label{fig:sz_labeled}
\end{figure}

To start, the groups were transformed to a singular ``CCI'' as previously explained.  The location of the hard and soft apexes were then defined by hand, and labeled with ranks 1 and 2 respectively.  Using the scale of the normal branch, ranks were then defined for the rest of the Z track.  The $s_z$ parameter was then obtained by spline interpolating the ranks with hard colour, soft colour, and intensity.  $t_z$ is easily calculated from there as the distance of the data to those splines.  Fig. \ref{fig:sz_labeled} shows selected $s_z$ values over the transformed data for all three groups.

Calculating $u_z$ is less straightforward than the other parameters.  With regular cylindrical coordinates, the angular and radial components are calculated on planes normal to the height axis, which means that the vectors defining the angular origins are parallel at any height.  For these parameterizations, the ``height'' axis is allowed to bend and twist in space, meaning that the normal planes do as well.  If the movement of these planes can be defined, along with an initial plane, then one can calculate the equation of a plane for any value of $s_z$, and thus the angular component.  Borrowing terminology from the similar Frenet-Serret formulae \citep{Frenet1852,Serret1851}, the tangent axis ($T$) is the one pointing along $s_z$, and the normal ($N$) and binormal ($B$) lie on the perpendicular plane.  The directions of $N_1$ and $B_1$ (vectors on the smallest $s_z$ plane) are arbitrary, so long as they are perpendicular.  To find $N_{k}$ and $B_{k}$ from $N_{k-1}$ and $B_{k-1}$, the $N_{k-1}B_{k-1}$ plane undergoes two separate rotations, which makes $T_{k-1}$ parallel to $T_{k}$ (see Fig. A2 in \citet{igl_scox1}).  First, $T_{k-1}$ is rotated around an axis that passes through the $T_{k-1}N_{k-1}B_{k-1}$ origin and is parallel to the intensity axis.  The resulting vector ($T_{k-1}'$) differs by an angle such that the projections of $T_{k-1}$ and the rotated $T_{k-1}'$ onto the soft-hard plane are parallel.  The second rotation is rotating $T_{k-1}'$ about its origin to be parallel with $T_{k}$.

It was decided that the aforementioned Frenet-Serret equations would not be used, even though they are applied to similar problems.  The reasoning comes from the definition of the $N$ axis, which is given as $dT/ds$, where $s$ is the arc length.  Thus, where $dT/ds$ changes direction (``wiggles''), the $N_{k}B_{k}$ plane could be rotated significantly when compared to the $N_{k-1}B_{k-1}$ plane.  In comparison, this system is reliant on a static axis (intensity), and so no extreme plane rotations should occur, unless $T_{k}$ passes through a vector parallel to the intensity axis.

\section{CCF Methodology and Reprocessing Search}
\label{sect:ccf}
CCFs taken of a set of real data are usually performed with either the interpolated CCF \citep{1987ApJS...65....1G} or the discrete CCF \citep{1988ApJ...333..646E}.  The DCCF is generally preferred, as data interpolated over larger gaps can dominate in a set of equally weighted interpolated points.  Gaps like this occur in cases such as the removal of bad data or clouds.  DCCFs take the form
\begin{equation}
    DCCF(\tau)=\frac{1}{M}\sum_{\tau-\Delta \tau/2}^{\tau+\Delta \tau/2}\frac{(a_i-\bar{a})(b_j-\bar{b})}{\sqrt{\sigma_a^2\sigma_b^2}}
\end{equation}
where $\tau$ is the lag, $a$ and $b$ are the data trains in question (X-ray and optical lightcurves in this paper), $\sigma_a$ and $\sigma_b$ are the standard deviations of the data trains, $M$ is the number of pairs per bin, and $\Delta \tau$ is the bin size.  For these reasons, all CCFs presented in this paper are discrete.

\begin{figure}
	\includegraphics[width=\columnwidth]{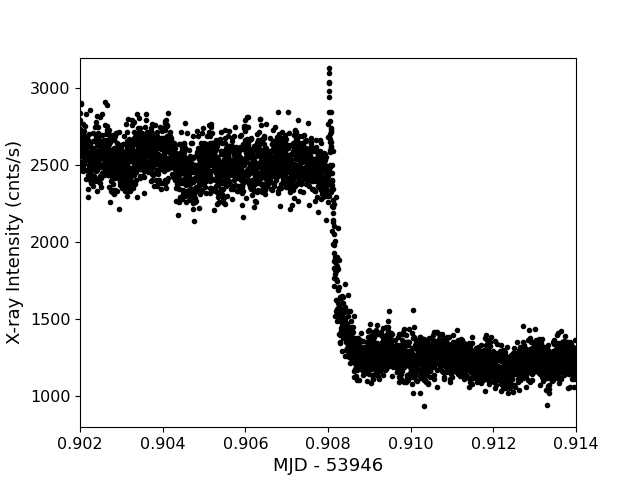}
    \caption{An example of the effect seen when the number of active RXTE PCUs changes.}
    \label{fig:pcu_switch}
\end{figure}

CCFs were obtained systematically by splitting the lightcurves into segments with no large gaps.  A CCF was then calculated for every instance of these split lightcurves overlapping in time.  Additional segments were created to split the lightcurves at changes in the number of active PCUs.  This was done instead of dividing the lightcurve by the active PCUs.  The main reason for this was an effect that can be seen in the lightcurves, where the transition to the new operating PCU number is not immediate, but instead lasts anywhere from 30 s to 2 min (Fig. \ref{fig:pcu_switch}).  The switch is sometimes accompanied by an intensity spike.  Thus, data within 2 minutes of the PCU change were considered ``bad'', and was not used in the CCFs.  Although this may seem a bit liberal, 4 minutes is negligible when compared to the amount of data in a lightcurve segment.

The timescale of a CCF feature (i.e. peak lag and width) is dependent on the physical processes within the binary that created it.  For reprocessing, X-rays produced near the central object can be reprocessed into optical wavelengths at distances as far as the companion star.  As such, the mark of potential reprocessing in LMXBs is a peak occurring at positive optical lags (X-rays lead the optical) on the order of seconds.  For Cyg X-2 specifically, these values can range from 0 s to $\sim$110 s.  Reprocessing on the companion follows the equation given in \citet{obrien_2000}:
\begin{equation}
    \tau=\frac{a}{c}-\frac{a}{c}sin(i)cos(\phi)
\end{equation}
where $a$ is the semi-major axis, $c$ is the speed of light, $i$ is the inclination, and $\phi$ is the binary phase.  Cyg X-2 has a quite high inclination ($62.5^\circ\pm4^\circ$ in \citet{1999MNRAS.305..132O}), meaning that reprocessing occurring on that region could cover a large range of lags ($\sim$5 -- 110 s).  Because much of the calculated CCFs are comprised of broad features, Butterworth filters (which have a maximally flat passband) with cutoff periods of 15 min were applied to the optical and X-ray lightcurves, to make potential reprocessing peaks (which would have much smaller lag widths) more visible and easier to identify.  The magnitude of the CCFs is much less important than the actual behaviors seen.

\begin{figure}
	\includegraphics[width=\columnwidth]{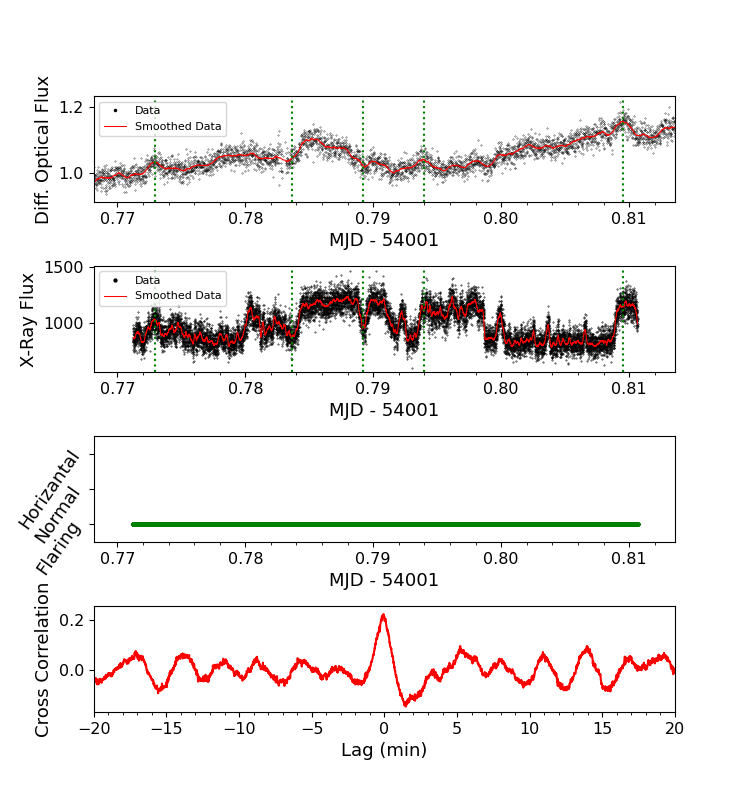}
    \caption{The CCF of MJD 54001 using data that has been Butterworth filtered.  The peak occurs at about -3 s of lag.  The dashed lines in the lightcurves occur at the same time for both the optical and the X-ray.  Note that 0.001 days corresponds to 1.44 minutes.}
    \label{fig:54001_ccf_filt}
\end{figure}

\begin{figure}
	\includegraphics[width=\columnwidth]{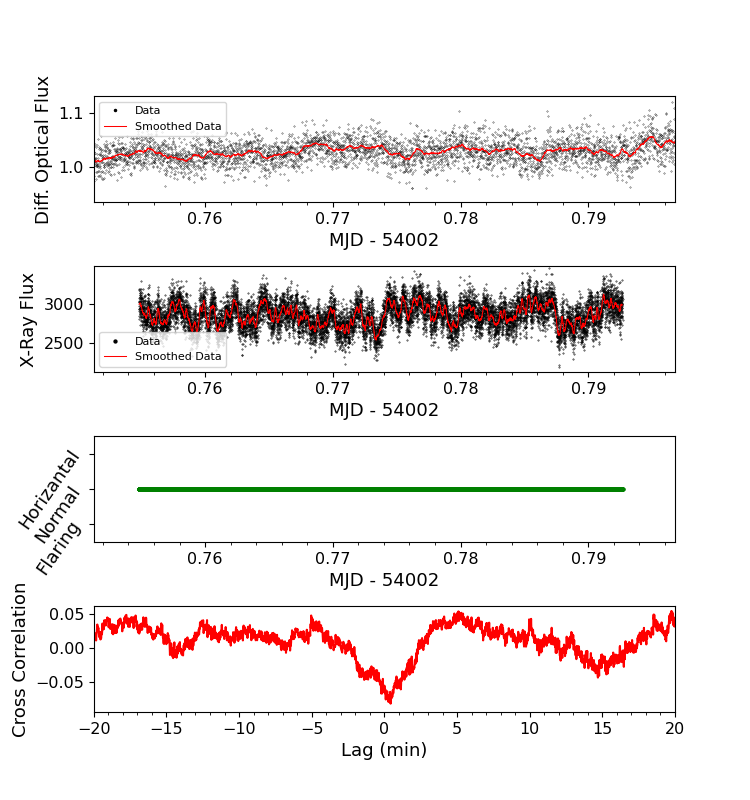}
    \caption{The CCF of MJD 54002, containing a clear anti-correlation at about 24 s of lag.  Note that 0.001 days corresponds to 1.44 minutes.}
    \label{fig:54002_ccf}
\end{figure}

\begin{figure}
	\includegraphics[width=\columnwidth]{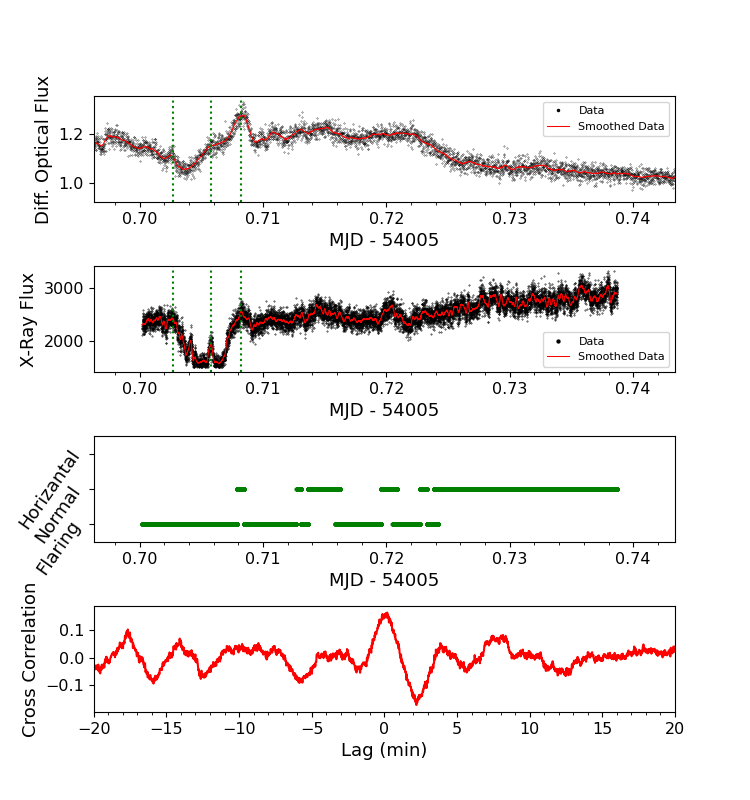}
    \caption{A CCF from MJD 54005 using data that has been Butterworth filtered.  The peak is centered at 0 s of lag.  Note that 0.001 days corresponds to 1.44 minutes.}
    \label{fig:54005a_ccf_filt}
\end{figure}

\begin{figure}
	\includegraphics[width=\columnwidth]{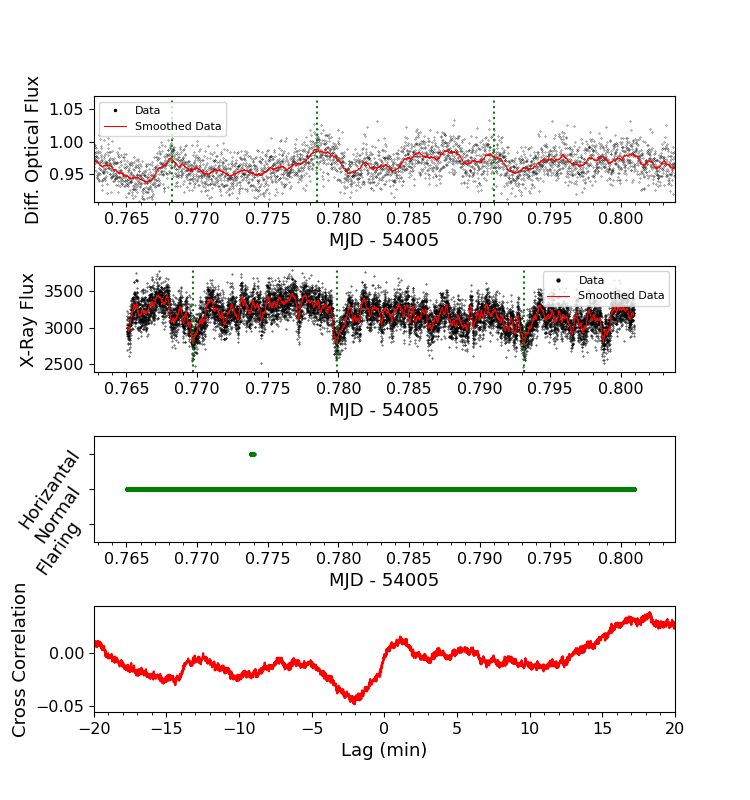}
    \caption{A CCF from MJD 54005.  The peak occurs at lags too high to be reprocessing, but the data contain multiple instances of optical peaks followed by X-ray dips.  Note that 0.001 days corresponds to 1.44 minutes.}
    \label{fig:54005b_ccf}
\end{figure}

Out of the entire dataset, four events of interest were identified, all of which were in group 3 and occurred at near-zero lag: one on MJD 54001, one on MJD 54002, and two on MJD 54005.  Most of these events also have corresponding features in their optical and X-ray lightcurves.  On its CCF, the MJD 54001 peak appears relatively small and difficult to see.  After applying the Butterworth filter to the data, the CCF peak becomes clear, standing out amongst the surrounding features (Fig. \ref{fig:54001_ccf_filt}). It has a large peak centered several seconds before a lag time of 0 s. This can be confirmed by checking the data themselves, where the close alignment of features is obvious. MJD 54002 has a possible event, but it is not likely that it is an echo (Fig. \ref{fig:54002_ccf}). Although it is well defined against the background and occurs at a lag on the order of seconds, the peak is negative (an anti-correlation). This is still an interesting occurrence, however, as Cyg X-2 is firmly on its NB, which is associated with X-ray/optical anti-correlations in Sco X-1 \citep{2016MNRAS.459.3596H}.  The first MJD 54005 plot has corresponding features which take place at nearly identical times, and as may be expected, the Butterworth filter reveals a large peak occurring at 0 s of lag (Fig. \ref{fig:54005a_ccf_filt}). As the X-ray time resolution is 1 s, this could be an instance of reprocessing from the inner disc. The second MJD 54005 plot has one peak that occurs at an optical lag of $\sim$1 min, which could suggest companion reprocessing, but the lack of peaks at similarly high lags makes this unlikely. However, it does contain some interesting behavior, in that there are multiple instances of rises in the optical, followed by drops in the X-ray (i.e., an anti-correlation at negative lag, Fig. \ref{fig:54005b_ccf}).

\begin{figure}
	\includegraphics[width=\columnwidth]{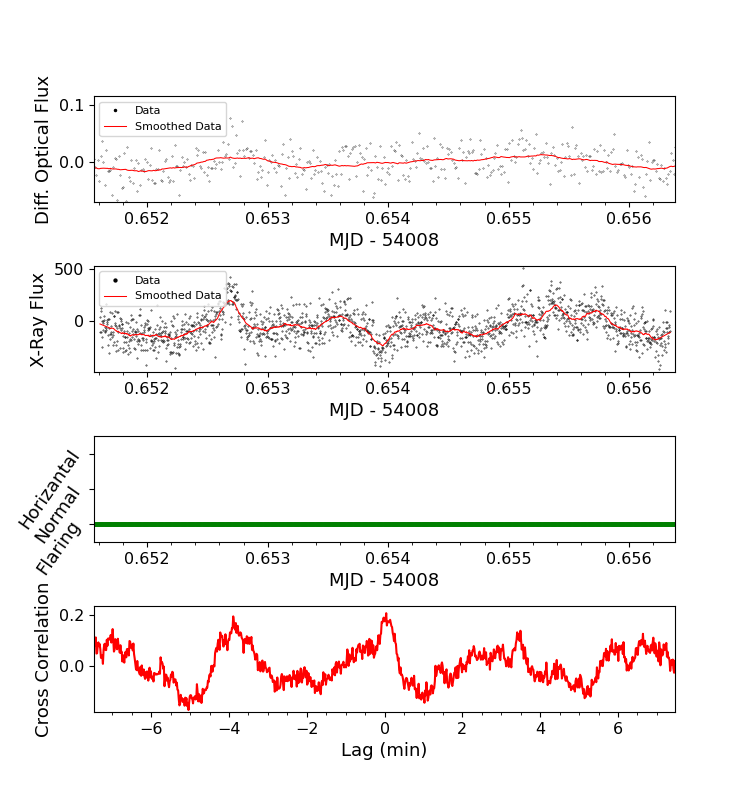}
    \caption{A CCF from MJD 54008, containing only FB data and high pass Butterworth filtered.  This reveals a thin CCF peak centered at $\sim$3 s.  Note that 0.001 days corresponds to 1.44 minutes.}
    \label{fig:54008_ccf}
\end{figure}

During the times investigated, Cyg X-2 was on all three Z source branches (Fig. \ref{fig:ccd}). Therefore, it was worthwhile to investigate how the object’s location on the CD affected the correlation between the optical and X-ray light curves.  No definitive trends were found, but the first MJD 54005 correlation (from Fig. \ref{fig:54005a_ccf_filt}) was dominated by the data in the flaring branch, as the peak disappears completely when the data is filtered for the NB, but remains when filtered for the FB. Another intriguing find was on MJD 54008. The cross correlation for the entire time period contains oscillatory behavior, with no clear peak that stands out among the others.  However, when selected for the flaring branch, a thin peak centered near zero lag ($\sim$3 s), as in Fig. \ref{fig:54008_ccf}. This was not repeated in any of the other Cyg X-2 data that were analyzed, but this near-zero lag peak structure can be seen in Figs. \ref{fig:54001_ccf_filt} and \ref{fig:54005a_ccf_filt}. Similar patterns were seen in the Z source Sco X-1, which saw positive peaks occurring only in FB and soft apex data \citep{igl_2023,igl_scox1}.  NB behaviors (Figs. \ref{fig:54002_ccf} and \ref{fig:54005b_ccf}) were also comparable, as \citet{2016MNRAS.459.3596H} noted broad anti-correlations in the same region.  Although there was a search for reprocessing during the times near the X-ray burst, no interesting features were found in the CCF.  This could be tied to the system being on the HB during that section of the lightcurve.

\section{Z Track Behaviors}
\label{sect:z_behave}
\begin{figure}
	\includegraphics[width=\columnwidth]{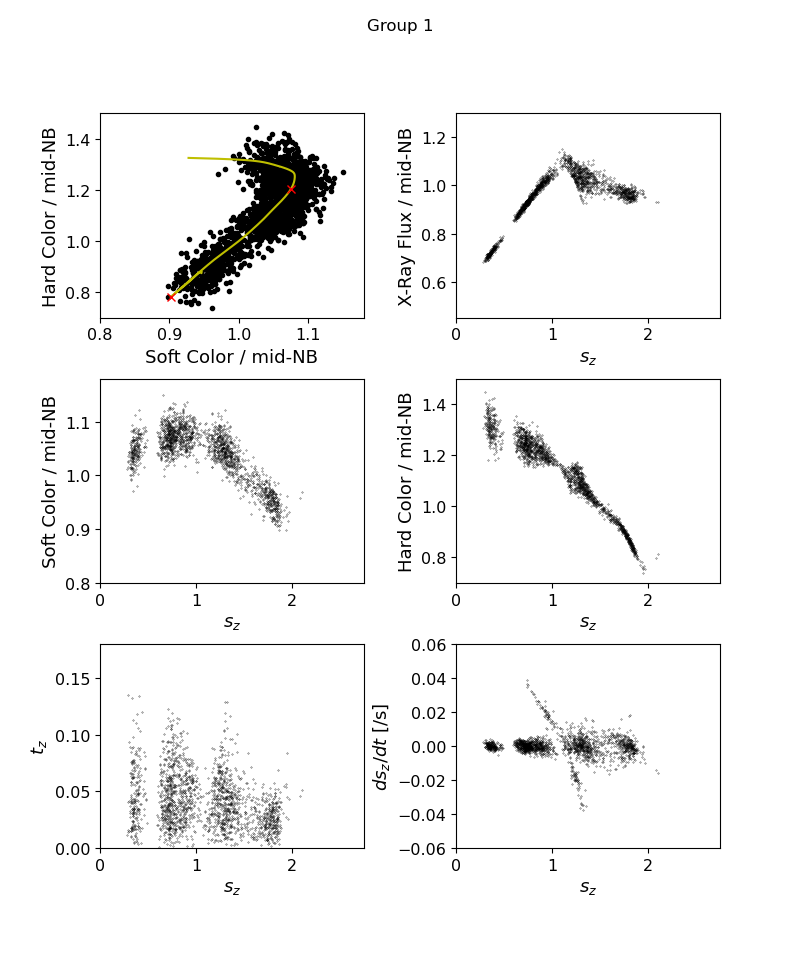}
    \caption{Various group 1 observables and parameters plotted against $s_z$.}
    \label{fig:g1_ranks}
\end{figure}

\begin{figure}
	\includegraphics[width=\columnwidth]{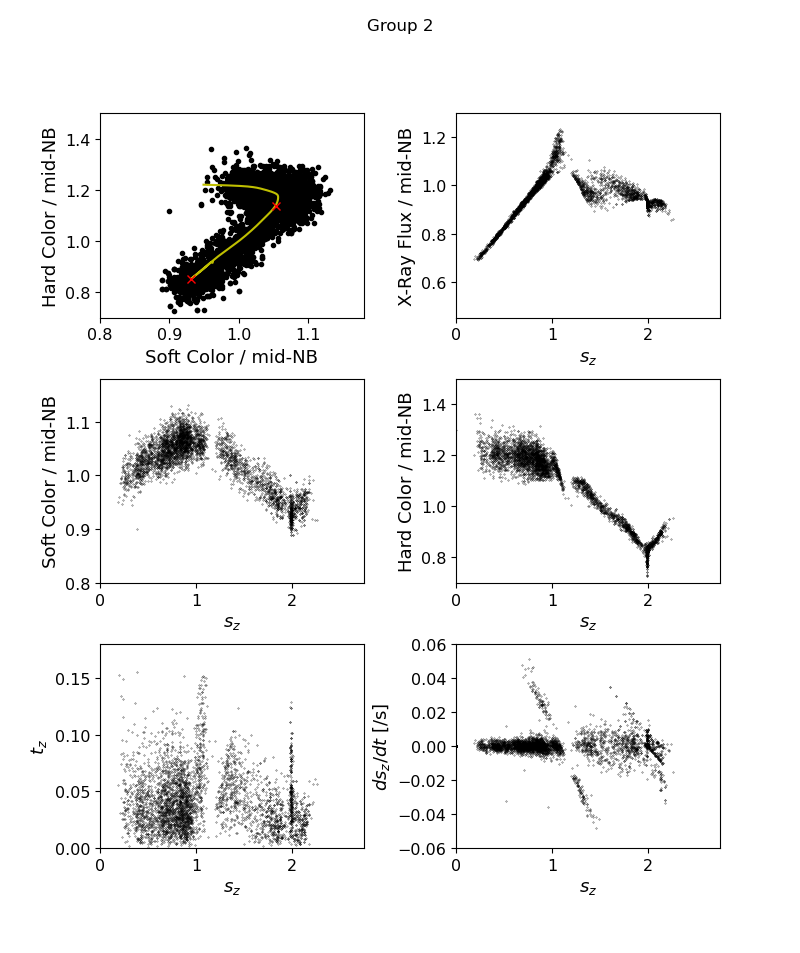}
    \caption{Various group 2 observables and parameters plotted against $s_z$.}
    \label{fig:g2_ranks}
\end{figure}

\begin{figure}
	\includegraphics[width=\columnwidth]{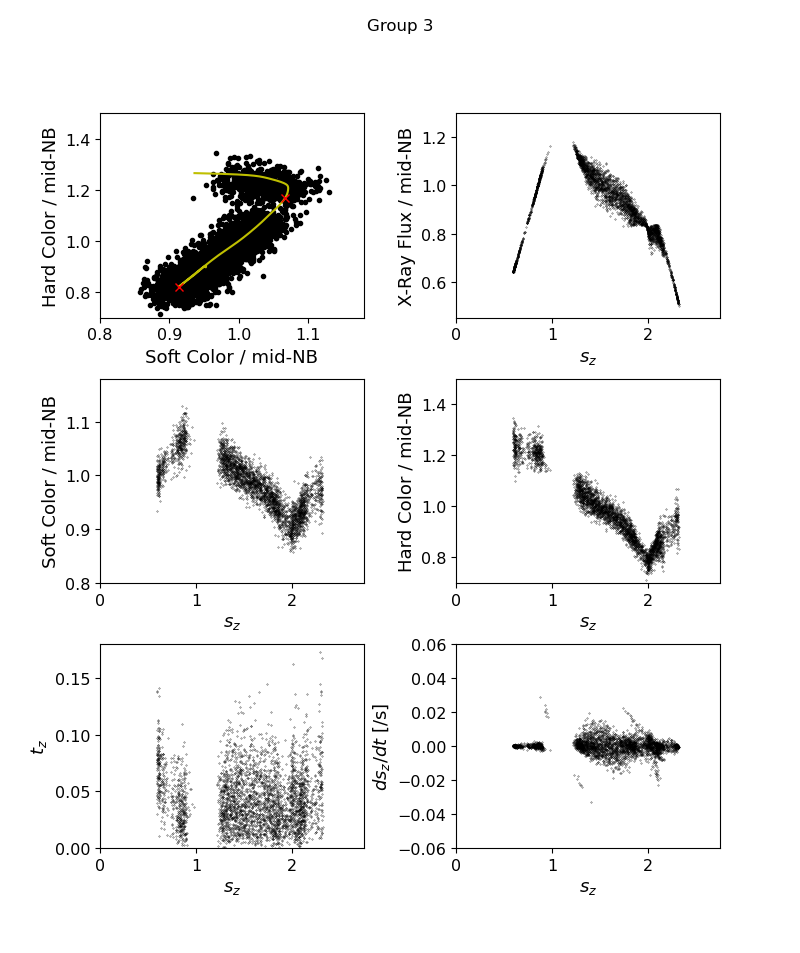}
    \caption{Various group 3 observables and parameters plotted against $s_z$.}
    \label{fig:g3_ranks}
\end{figure}

The results of the ($s_z$, $t_z$) transformations can be seen in Figs. \ref{fig:g1_ranks}, \ref{fig:g2_ranks}, and \ref{fig:g3_ranks} for groups 1-3 respectively. For all three groups, the soft, hard, and flux trends as a function of $s_z$ are approximately linear, changing slope at the apexes.  There are some deviations from this, however, occurring most obviously towards the middle of the NB.  The $s_z$ vs. $t_z$ plot also makes clear that the CCI plot is quite consistent through $s_z$ in terms of radial thickness, save for a broadening near the hard apex ($s_z$=1).

The time derivative of $s_z$ shows different behaviors for each group. \citet{1992ApJ...396..201H} and \citet{2000MNRAS.311..201D} both found a clear taper in $ds_z/dt$ at the soft apex of Sco X-1, but this does not appear present in our Cyg X-2 data. Group 2 (Fig. \ref{fig:g2_ranks}) may have a taper at the hard apex, which is the opposite of the results in \citet{2000MNRAS.311..201D}, where a local maximum of the speed along the Z track is reached.  Note that both \citet{1992ApJ...396..201H} and \citet{2000MNRAS.311..201D} calculated $s_z$ using only soft and hard colours.  The inconsistencies in the $s_z$ definitions may account for the differences in these behaviors.

In addition, there is a noticeable increase in scatter of the time derivative in the NB for all three groups. The noise component in these measurements is likely small, as indicated by the low scatter in the HB section of Fig. \ref{fig:g3_ranks} as compared to the NB.  Another common thread for the groups, but especially group 2, is a region of drastically higher and lower $ds_z/dt$ values centered at the apexes.  On the concave side of two branches, the mapping tends to push consecutive points farther away from each other, and if these points are mapped to separate branches, $|\Delta s_z|$ can become quite large.  Therefore, even if the distances between consecutive data points are similar, the magnitude of $ds_z/dt$ = $\Delta s_z/\Delta t$ can be outsized in these regions.  Large positive values appear at lower $s_z$, as the system begins moving toward the FB, and large negative values appear at higher $s_z$, as the system moves toward the HB.  Thus, these sharply-sloped features are not due to binary behavior, but are artifacts purely a result of mapping.

\begin{figure}
	\includegraphics[width=\columnwidth]{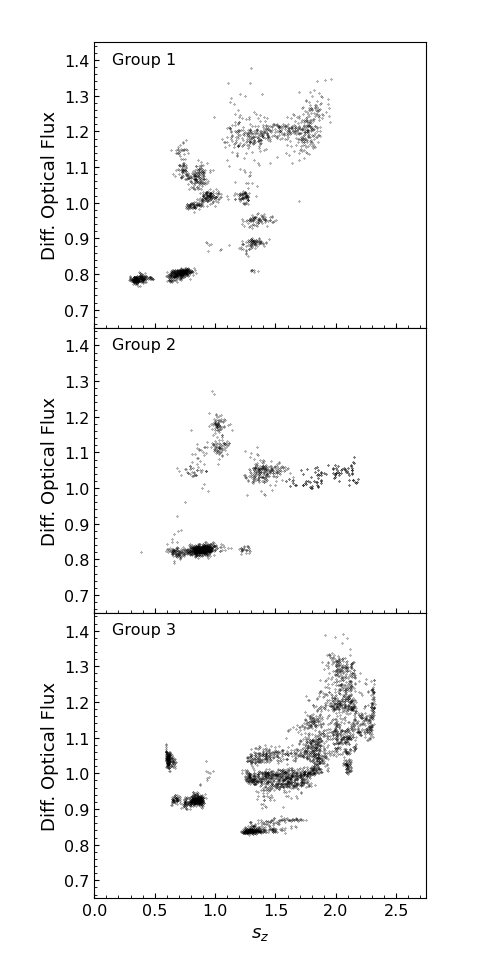}
    \caption{Plots of optical intensity against $s_z$ for all three groups.}
    \label{fig:sz_opt}
\end{figure}

Fig. \ref{fig:sz_opt} shows the Cyg X-2 optical data plotted against $s_z$. The most immediately interesting behavior is in the top plot (group 1), which takes somewhat of an ``S'' shape. This carries the implication that, as the optical flux decreases, Cyg X-2 is bouncing between the HB and NB. It also makes clear that the optical flux is multi-valued, and cannot be uniquely determined by $s_z$. The bottom plot (group 3) may show this as well, with a switch to the HB at a differential optical flux of about 0.95. Group 3 extends further into the FB than group 2, and there appears to be a positive correlation where $s_z>2$. In the HB and NB, the optical data remain multi-valued. Group 2 looks very different from groups 1 and 3, in that there seems to be no correlation of $s_z$ and the optical flux. Thus, the data suggest that there are two different relations to optical behavior along the Z track (a no-correlation state and a multi-valued/correlation state).  It is worth noting that, along with its optical-$s_z$ plot, group 3 has a HID that looks quite different than the other two (Fig. \ref{fig:hid}), containing both a FB and a noticeably different NB slope.

\begin{figure}
	\includegraphics[width=\columnwidth]{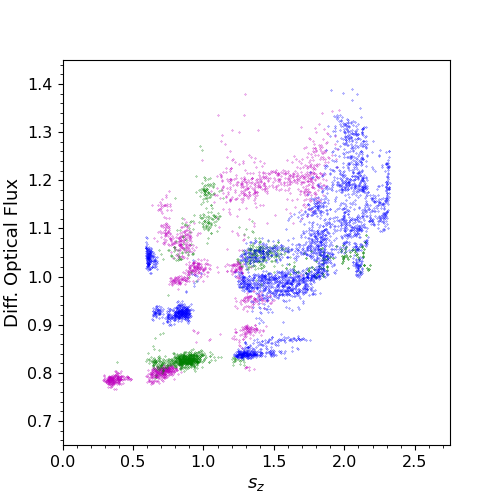}
    \caption{The optical-intensity plot of all three groups.}
    \label{fig:sz_opt_combined}
\end{figure}

Because secular drift is accounted for when calculating the Z track parameters, all three plots in Fig. \ref{fig:sz_opt} can be combined, with the groups directly compared (Fig. \ref{fig:sz_opt_combined}).  Viewed like this, it looks as though the HB and NB are comprised of steps.  These steps have very little optical scatter by themselves, but the optical locations of each step can occur at a wide variety of values.  The collated plot reframes the group 2 data as well.  Instead of being a completely uncorrelated state, it now appears as more HB and NB steps, along with a potential transition on the hard apex.  Similar transitions have been seen in Sco X-1 (\citeauthor{igl_scox1}, submitted).  The difference between group 2 and groups 1 and 3 might then simply be that group 2 is less well-sampled.

\begin{figure}
	\includegraphics[width=\columnwidth]{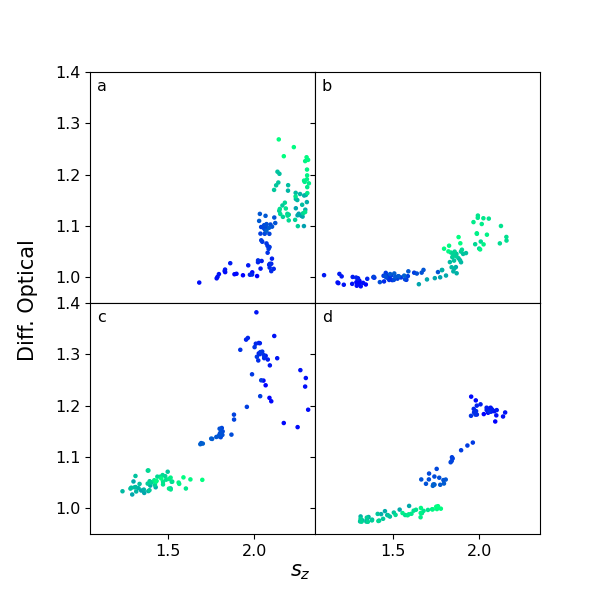}
    \caption{Optical-$s_z$ plots for four separate group 3 observations.  Colours move from blue to green for earlier and later times respectively (normalized to each observation).}
    \label{fig:steps}
\end{figure}

Looking at the group 3 (the only group with FB data) $s_z$-optical plots by observation reveals another interesting trend: evidence of a clear jump in the optical near the soft apex (Fig. \ref{fig:steps}), moving from a low scatter NB level to a high scatter FB level.  Similar behavior has been noted in Scorpius X-1, which included a step at the hard apex as well (\citeauthor{igl_scox1}, submitted).  Note that even though the jumps do not always occur exactly at $s_z$=2, it is likely that the soft apex is moving even within the group, as Cyg X-2 is known to have particularly high secular drift.  The behavior of the jump depends on the direction in $s_z$ the system is heading.  When moving to higher $s_z$, the jump is more like a step, moving directly up to a new optical level, before continuing onto the FB with higher variability than was seen on the NB.  With descending $s_z$, the jump is a slope, descending linearly down to the new optical flux in $\delta s_z\simeq0.3$.  Note that although group 3 has an apparently large scatter in the NB (comparable to the scatter in the soft apex and FB), this occurs over many observations.  Single observation scatter is much smaller, but the flux level itself varies.  This reframes the $s_z$-optical correlation in group 3 and its scatter (days worth of data) as a combination of several slopes and steps from repeated crossings of the soft apex on the order of hours.  

\section{Spectral Fitting}
\label{sect:spectral}
Data in the energy range of 3-20 keV were passed into \textit{Xspec} (version 12.12.0) for spectral fitting.  A number of models commonly applied to Z sources were tested using spectra from a full observation of data ($\sim$1-2 hours), but none were found to be a clear best fit.  The adopted model was similar to the well known Extended ADC model \citep{1995A&A...300..441C,2010A&A...512A...9B}, TBABS(BB+CPL+GAUSS), where TBABS is an absorption model, BB is a blackbody (soft X-ray source), CPL is a cutoff power law (hard X-ray source), and GAUSS is a Gaussian to represent the iron K$\alpha$ line.

\begin{figure}
	\includegraphics[width=\columnwidth]{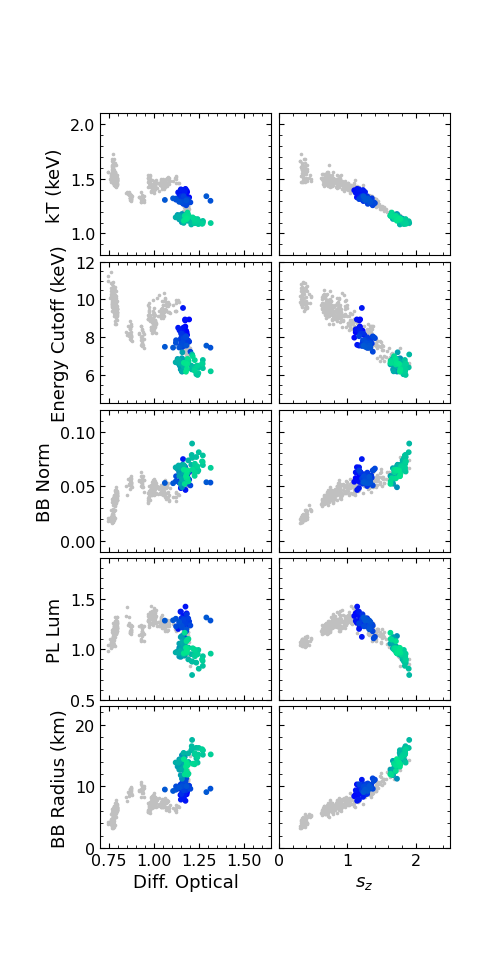}
    \caption{Group 1 spectral fits.  The highlighted points are from the same observation (blue represents earlier times, green represents later times), and show how the correlated behavior in the high optical occur at the soft apex.  The blackbody normalization is in units of 10$^{39}$ (ergs/s)/(10 kpc)$^2$, and the cutoff powerlaw luminosity is in units of 10$^{38}$ ergs/s.}
    \label{fig:g1_specs}
\end{figure}

\begin{figure}
	\includegraphics[width=\columnwidth]{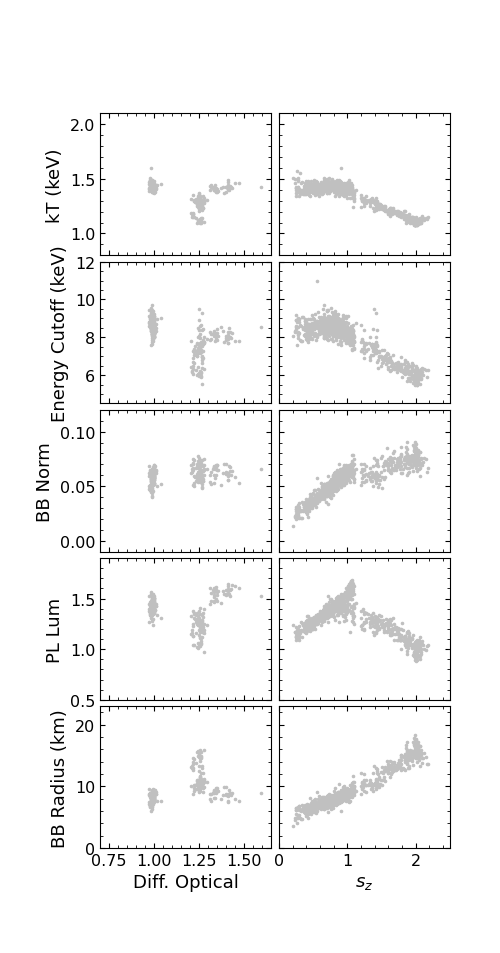}
    \caption{Group 2 spectral fits.  The blackbody normalization is in units of 10$^{39}$ (ergs/s)/(10 kpc)$^2$, and the cutoff powerlaw luminosity is in units of 10$^{38}$ ergs/s.  No points are highlighted here because there are no distinct optical correlations.}
    \label{fig:g2_specs}
\end{figure}

\begin{figure}
	\includegraphics[width=\columnwidth]{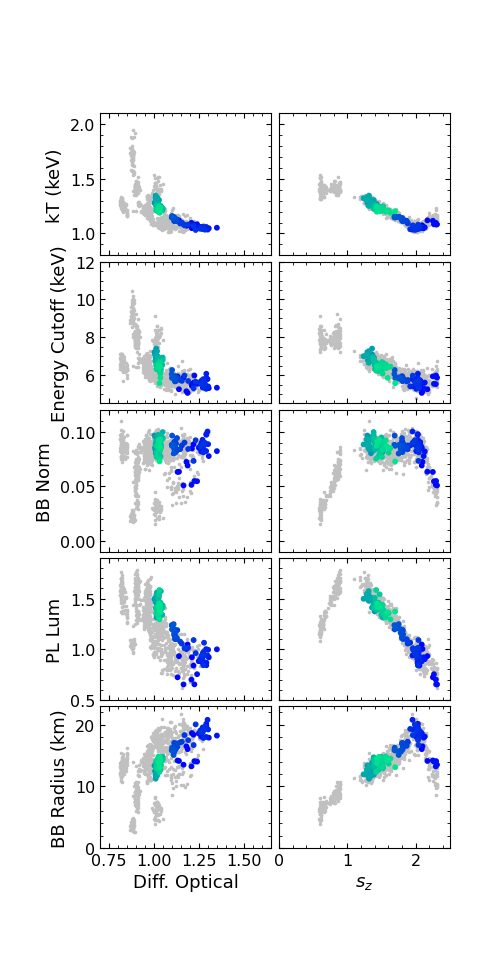}
    \caption{Group 3 spectral fits.  The highlighted points are from the same observation (blue represents earlier times, green represents later times), and show how the correlated behavior in the high optical occur at the soft apex.  The blackbody normalization is in units of 10$^{39}$ (ergs/s)/(10 kpc)$^2$, and the cutoff powerlaw luminosity is in units of 10$^{38}$ ergs/s.}
    \label{fig:g3_specs_d}
\end{figure}

Once the model had been chosen, a fitting strategy needed to be selected in order to produce an accurate fit.  This involved testing fitting parameters in different orders, as well as freezing parameters at their initial values for the entirety of fitting.  Plotting the fitted parameters versus time was the method used to check if the fits were sound.  A good fit was free from parameters that hit a hard minimum/maximum, as well as instances where multiple parameters drastically changed without reason within the data.  Two consecutive fits were used to obtain the parameters from a full observation of data: (1) a fit where the Gaussian parameters are frozen, and (2) a fit where every parameter except for the Gaussian is frozen.  The absorption coefficient was frozen for both fits at $0.2\times10^{22}$ cm$^{-2}$.  Although locking $N_H$ is not ideal, as it can vary, fits could not be stably constrained otherwise.  Other examples exist in the literature where $N_H$ is frozen as well \citep{2007ApJ...667..404T,2021NewA...8301479D}.  In addition, as is commonly done in other works fitting XRBs \citep{2000ApJ...533..329B,2015JApA...36..335D}, a systematic uncertainty of 0.5\% was added to the fits.

The time-resolved spectra required more finesse to obtain a good fit, so a slightly different process was used.  Firstly, because the K$\alpha$ lines proved particularly difficult to fit, it was decided that the Gaussian energies for the higher time resolution spectral fits would be frozen at the values from the full observation (hours of data) spectral fits.  The absorption parameter $N_H$ was again frozen to 0.2$\times$10$^{22}$ cm$^2$.  The value of the cutoff power law index was also frozen at a value of 1.7, as \citet{2010A&A...512A...9B} found that the parameter remained close to 1.7 when fit, for Cyg X-2 and other Cyg-like sources \citep{2006A&A...460..233C}.  The power law cutoff energy was temporarily frozen at its value from the full observation fit.  After an initial fit, the cutoff energy was freed, and the model was refit to give the final values.

This process was performed on data reduced at 16, 32, 64, and 128 second intervals, so that a balance between fit noisiness and a clear idea of how the fitting parameters were behaving in time could be found.  Plots of parameter fits vs. time for the 16 second spectra were extremely noisy, making it clear that a lower time resolution was necessary.  The noise levels in the 32 s spectra decreased significantly, and even more so in the 64 s spectra.  Ultimately, it was decided that the 64 s spectra would be used, because on plotting certain parameters vs. optical data, the additional time resolution in the 32 s spectra did not seem to add any new information to the plots.

Results of the timed spectra fitting can be seen in Figs. \ref{fig:g1_specs}, \ref{fig:g2_specs}, and \ref{fig:g3_specs_d} (more group 3 fits can be found in \citet{igl_2023}), plotted against $s_z$ and differential optical flux.  The $s_z$ plots show clearly that most of the physical parameters behave monotonically within a branch.  $kT_{BB}$ and the powerlaw cutoff energy vs. $s_z$ both show very similar behaviors: near constant values in the HB, a decrease moving from hard to soft apex, and an increase from soft apex through the FB.  In all three groups, the $kT_{BB}$ plot has a noticeable increase in scatter at the lowest $s_z$ values.  The same parameter behaviors can be seen in \citet{1990A&A...235..131H} when fitting with a blackbody and a Boltzmann-Wien spectrum.

The blackbody normalization, powerlaw luminosity, and blackbody radius all follow different patterns.  Note that the blackbody radius was calculated using the equation
\begin{equation}
    R=\sqrt{\frac{L}{4\pi\sigma T^4}}
\end{equation}
where $L$ is the blackbody luminosity (obtained from the blackbody normalization), $\sigma$ is the Stefan-Boltzmann constant, and $T$ is the blackbody temperature.  All three parameters monotonically increase moving through the HB to the hard apex.  Continuing onto the NB, the blackbody normalization and radius continue to increase, albeit not as quickly for the normalization (the parameters are barely correlated in group 3).  This results in the radius rate of increase to be the same as in the HB, despite the change in both the blackbody normalization and temperature.  Group 1 (Fig. \ref{fig:g1_specs}) is an exception to this, as starting in the mid-NB, the rate of increase changes to a new value for both the blackbody normalization and the powerlaw luminosity.  The $s_z$ where this begins ($\sim$1.6) is likely too low to attribute to secular drift of the soft apex.  Of the three aforementioned parameters, the powerlaw luminosity is the only one that begins to decrease in the NB.  The rate of NB decrease appears somewhat dependent on the rate of increase in the HB, with group 3 showing steep slopes in both, while the correlation decreases in groups 1 and 2.  In the FB, all three parameters decrease with $s_z$, with the group 3 powerlaw luminosity decrease remaining relatively unchanged from its behavior in the NB.

The optical-fitted parameter plots in Figs. \ref{fig:g1_specs}, \ref{fig:g2_specs}, and \ref{fig:g3_specs_d} display distinct behaviors for each group.  Group 1 has a defined ``wiggle'' in all of the parameters, similar to the group 1 optical-$s_z$ plot in Fig. \ref{fig:sz_opt}, centered near $s_z=1$ (Fig. \ref{fig:g1_specs}).  It also contains a correlation between the optical and blackbody radius at the highest optical fluxes.  $kT$ and the cutoff energy have a maximum at the lowest optical intensities and minimum at the highest, and vice versa for the blackbody normalization and blackbody radius.  Group 2 contains no correlation at all between the optical flux and fitted parameters, which one may expect given the decoupling of the optical and X-ray intensities (Fig. \ref{fig:g2_specs}).  Group 3 shows no correlation between the optical and fitted parameters when the differential optical flux is less than 1.0, accompanied by a high optical scatter (Fig. \ref{fig:g3_specs_d}).  Higher optical intensities contain clear correlations with blackbody parameters (normalization and temperature, and consequently blackbody radius, see highlighted data in Fig. \ref{fig:g3_specs_d} and additional group 3 fits in \citet{igl_2023}).  Ultimately, it does not appear as though the optical correlates any better with these physical parameters than it does with $s_z$.

The exception to the above statement may lie with the powerlaw luminosity, which shows a decrease in the parameter with increasing optical.  The behavior is seen most clearly in group 3, where the luminosity hovers around 1.5$\times$10$^{38}$ ergs/s, and dips down to about 0.8$\times$10$^{38}$ ergs/s at a differential optical value of 1.0.  This dip is clearly present in group 1 as well, although the disparity between luminosity steps is not as large, and the change occurs at a higher optical value.  The optical coverage in group 2 is too low to make a conclusive statement about the behavior, but the powerlaw luminosity drops to a lower level somewhere between 1.0 and 1.25 on the optical, and jumps to higher levels when the system is optically brightest.  The jump may imply that the pattern is inconsistent with the other groups.

\section{Discussion}
\label{sect:discussion}
\subsection{Cross Correlation Behaviors}
\citet{2015MNRAS.451.3857S}, who observe optical-XR anticorrelations on the NB and near-zero positive correlations on the FB, discuss and interpret multiple models that can explain this in our CCFs.  In the \citet{1995ApJ...454L.137P} model, soft photons are emitted by the neutron star magnetosphere, and hard photons are emitted when the soft ones are Comptonized in a hot central corona or the radial inflow from an outer corona.  In the NB, the accretion rate increases, leading to higher radiation pressure and a pileup of material around the neutron star.  X-ray photons are then absorbed and reemitted in the optical regime by this material.  The more photons are absorbed, the fewer reach the observer in the X-ray regime, which would lead to an anti-correlation.  On the FB, the electron scattering optical depth has increased enough that X-ray light is being scattered onto the outer disc or companion, leading to reprocessing on those regions.  An increase in X-ray intensity leads to more photons being scattered and reprocessed (and vice versa), resulting in an optical lag and a positive CCF peak.

In the extended ADC model from \citet{2010A&A...512A...9B}, the NB is also associated with increasing accretion rate, although this time it is when moving away from the soft apex.  \citet{2015MNRAS.451.3857S} predicts that this would still lead to anti-correlations due to the optical depth increasing with the accretion rate.  Moving up the FB, the extended ADC model predicts that Cyg-likes will have a constant accretion rate, but an increasing blackbody luminosity.  Thus, FB reprocessing still fits within the framework of the extended ADC model.  However, these data show that the blackbody luminosity is decreasing through the FB (Fig. \ref{fig:g3_specs_d}).  It is possible that the reason for this lies in the distinction between a ``dipping'' branch and a ``flaring'' branch, a more in depth discussion of which is contained in Section \ref{sect:phys_behav}.

Due to the short lags of the potential reprocessing peak maxima, it is likely that most of the reprocessing would be taking place on the accretion disc.  Even though Cyg X-2 has a large orbit and a high inclination (meaning that companion reprocessing lags could be seen as low as $\sim$6 s), the largest and most obvious CCF peaks occur at optical lags of around 0 s.  Fig. \ref{fig:54005b_ccf} has a small peak that occurs at about 1 min of lag, but being the only one in the tens of seconds range within this comprehensive dataset makes it an unlikely candidate for companion reprocessing.  The lags seen in these data are similar to the lags observed in Sco X-1 \citep{igl_scox1}, in spite of the different orbital periods.

In the \citet{igl_scox1} Sco X-1 dataset, a number of ``minor peaks'' appeared in the shape of well defined piecewise exponential functions at optical lags of less than 4 s.  These peaks were very small compared to surrounding CCF features, and were likely weaker versions of more obvious reprocessing peaks.  After a search, it was determined that minor peaks do not appear in these data.  The few possible reprocessing events that appear are not necessarily comparable in width or shape (i.e. Figs. \ref{fig:54001_ccf_filt} and \ref{fig:54008_ccf}).  The minor peaks were also much more common, with eight appearing over the course of the nine nights with overlapping optical and X-ray data.  However, in both datasets, reprocessing events occurred on the FB or soft apex, with none appearing outside of those regions.

The question could be raised why so little solid evidence of reprocessing is seen within this dataset, especially in the secondary star.  \citet{1999MNRAS.305..132O} argues that heating of the companion star is not a large contribution based on the small amount of excess light at the photometric phase 0.5.  Their model has the edge of the disc shielding the companion from the central X-ray source.  They also note that the disc is fainter than the companion in the optical (a factor of 9 smaller than anticipated, based on \citet{1994A&A...290..133V}), which might be unexpected, as one would think the disc would receive more illumination.  One possible solution they provide is that the X-rays are reprocessed into the ultraviolet, a regime which this comprehensive study does not attempt to detect.

Also of note is the unusually long period of Cyg X-2, $\sim$9.8 days for one full orbit.  This naturally leads to a large separation between the X-ray source and the companion, as well as an accretion disc with a bigger radius.  As the donor heating drops off with inverse distance squared, one may expect that it would be more difficult to see reprocessing in Cyg X-2 because the X-ray illumination is less intense and leads to less heating compared to the intrinsic luminosity of the A9 companion.

\subsection{Z Track Behaviors}
\begin{figure}
	\includegraphics[width=\columnwidth]{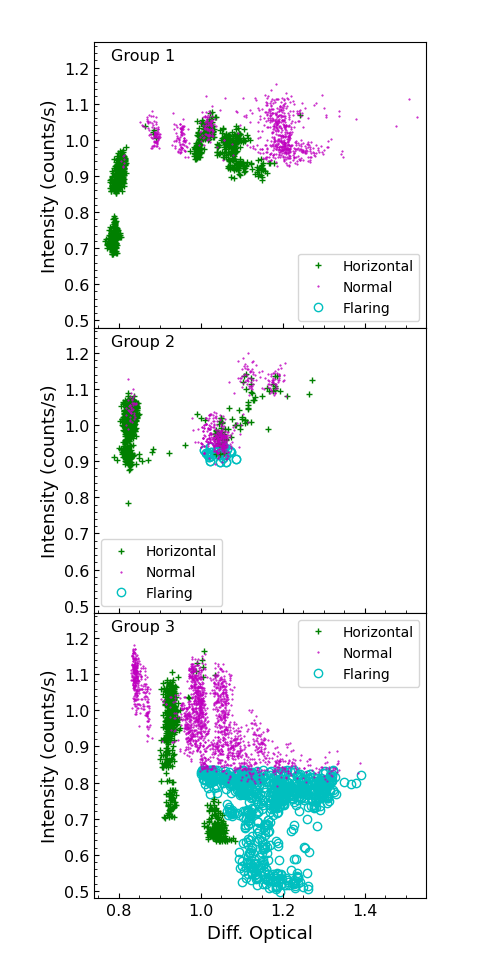}
    \caption{Optical-X-ray intensity plots for all three groups.}
    \label{fig:oi}
\end{figure}

These results can be contrasted with the work of \citet{2004MNRAS.350..587O}, who observed that (for Cyg X-2) optical flux tended to monotonically increase from HB to FB.  They also found that X-ray and optical intensities did not have a simple one-to-one relationship, and that the data could be contained within an envelope comprised of three spectral components, described by the equation
\begin{equation}
    I_X(F_\nu)=\alpha(t)I_F(F_\nu)+I_B(F_\nu)+I_A(F_\nu)
\end{equation}
Here, $I_X$ is X-ray intensity, $F_\nu$ is the optical flux, $I_F$, $I_B$, and $I_A$ are the flaring, baseline, and accretion spectral components respectively, and $\alpha$ is a coefficient that describes variability in $I_F$, such that $0\leq\alpha(t)\leq1$.  The group 1 data (Fig. \ref{fig:oi}, top plot) follows the scheme laid out in this equation quite nicely.  Because group 1 does not stretch into the FB, not much can be said about the baseline behavior.  However, the value of $I_F$ seems to be increasing with $s_z$.  \citet{2004MNRAS.350..587O} interprets this behavior as a result of inhomogeneities in the accretion flow, resulting in more variability in higher accretion rate (higher $s_z$) branches.  The group 2 plot contains less data, but HB and NB data within could conceivably be following the same pattern as in group 1, with similar conclusions about the spectral components.  Group 3 looks much different than the other two, with a more significant X-ray decrease during the NB.  Here, $I_F$ is larger than in both groups 1 and 2 at high optical fluxes, and remains large even through most of the envelope.  The group 1 and 3 plots in Fig. \ref{fig:sz_opt} also both support the idea that the $s_z$ parameter is more useful than X-ray flux at tracing optical behavior.  This is because, although accompanied by multi-valued behavior at low $s_z$, the optical has a general increasing monotonic trend across the Z track (for two different Cyg X-2 data groups).

The optical-$s_z$ states show very distinct behavior between the groups, and in group 2, the optical flux appears to be completely decoupled from the $s_z$ parameter (although this could be due to the data sampling).  Figs. \ref{fig:sz_opt} and \ref{fig:oi} falsify the null hypothesis that the optical perfectly traces X-ray radiation.  One possible explanation of this could be disc warping, as in \citet{1997MNRAS.292..136P}, where X-ray illumination from the central object causes a growth in the form of a prograde spiral on the inner accretion disc (where the X-rays are strongest).  Such distortions to the disc are often used to explain super-orbital periodicities, including with Cyg X-2 \citep{2003PASP..115.1124V}.  A warp of this nature could allow for outer disc illumination in some regions, and shielding in others, preventing reprocessing and decoupling the X-ray and optical behavior in these sections.  The amount of reprocessed light that reaches the observer would be dependent on the inclination of the system and the precession phase of the warp.  This scenerio would be consistent with the lack of reprocessing events found in the group 2 data.

Ultimately, XR intensity is better correlated with the Z track location than the optical flux.  Fig. \ref{fig:oi} shows that no single XR-optical intensity relation applies to the system at any given time, as while the group 1 data resembles what is seen in \citet{2004MNRAS.350..587O}, group 3 is quite different.  Thus, the optical is likely not as useful for understanding the system and constraining models.

\subsection{Physical Behaviors}
\label{sect:phys_behav}
In the \citet{2010A&A...512A...9B} extended ADC model, the mass accretion rate is at a minimum at the soft apex ($s_z=2$).  It increases as $s_z$ decreases towards $s_z=1$ (the hard apex), along with a decreasing blackbody radius, which can be interpreted as the emitting region shrinking down to an equatorial strip.  This, combined with the flux reaching super-Eddington values at the hard apex, causes enough radiation pressure to disrupt the inner disc and move matter vertically to form jets in the HB.   On the FB, the blackbody luminosity increases while the ADC luminosity remains constant, which implies that the neutron star has gained a non-accretion powered energy source.  The accretion rate divided by the blackbody area ($\dot{M}/4\pi R^2_{BB}$) drops lower than the critical value for unstable burning, leading to flaring.  Thus, in the FB, the increase of the blackbody luminosity was interpreted as power from nuclear burning with constant ADC luminosity (constant accretion rate).  Note that the ``Z'' shaped HID in \citet{2010A&A...512A...9B} looks different than the more common ``C'' shaped HID seen in these data.

$kT$ and $E_{CO}$ behave very similarly to what is seen in \citet{2010A&A...512A...9B}, whose observations also show little to no decrease through the HB, a greater reduction in the NB, and increasing again in the FB.  The authors posit that the $kT$ minimum at the soft apex corresponds to an accretion rate minimum at the same spectral location.  The blackbody radius here is at a maximum, which would mean that the entire neutron star is emitting.

\begin{figure}
	\includegraphics[width=\columnwidth]{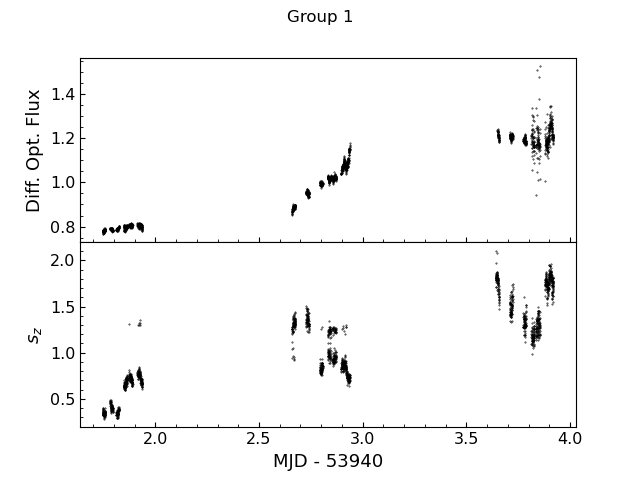}
    \caption{The group 1 optical and $s_z$ data plotted against time.}
    \label{fig:time_sz}
\end{figure}

At the hard apex, \citet{2010A&A...512A...9B} also found that the blackbody radius decreased, and continued to do so into the HB.  They interpret this as the region of blackbody emission shrinking down to an equatorial strip, which then has an increased emissive flux.  The greater radiation pressure moves accreted material from the inner disc vertically, potentially forming jets.  Our results show that this inner disc disruption may undergo a period of ``uncertainty'' before fully returning to the NB.  Fig. \ref{fig:time_sz} shows the system bouncing between the HB and NB before continuing towards the FB in group 1.  However, even while Cyg X-2 moves around the hard apex, the optical continues to increase.  This behavior was not found in group 3, where the hard apex optical level does not have any general trend with time.

Ultimately, the results of the spectral fitting reveal that the Z track location is a far better way to track physical behaviors than the optical flux in Cyg X-2 data.  Although interesting trends can be found within the optical plots, such as correlations during periods of high optical intensities, the $s_z$ plots display clearer positive and negative correlations by branch.  In addition, the behaviors of these parameters are unique with regards to Z track location.  Other than the cutoff energy and the blackbody temperature, each parameter has a different sequence of positive and negative correlations by branch.  The optical plots, however, have more similarities between parameters, and lack consistency in patterns between groups, with the relationship to physical parameters often being multi-valued.

\section{Conclusions}
\label{sect:conclusions}
In this study, we analyzed simultaneous optical (Argos) and X-ray (RXTE) lightcurves with 1 s time resolution.  Performing discrete cross correlations on overlapping segments of optical and X-ray lightcurves revealed both positive and negative correlation peaks occurring near zero lag in the CCF, many of which had visible corresponding features in the lightcurves.  The clearest near zero positive peaks all occurred on the FB, where reprocessing is normally seen.  Only one instance of an anti-correlation was seen, but it occurred while Cyg X-2 was on the NB, which is associated with such features.  Filtering data by branch further solidified the aforementioned conclusions.  Potential reprocessing peaks that were present with all of the data disappeared when looking only at the normal branch data, and peaks that did not exist in the CCFs using all of the data in a segment appeared when using only the FB data.

The Z track was parameterized using a modification of the rank number scheme.  Instead of performing spline interpolations on only the hard and soft colours, the X-ray intensity was included to account for the doubling back of the FB onto the NB in the group 3 data.  Of the most interest was the $s_z$-optical plots, which showed two different behaviors:
\begin{itemize}
    \item A multi-valued/correlated state:  This can be seen in the group 1 and 3 data.  The Z track position does not uniquely determine the optical intensity in the HB and NB.  In the FB, the two are correlated.
    \item A no correlation state: In group 2, there is no correlation between optical intensity and Z track position on any branch.  However, this effect may be due to the limited amount of data.
\end{itemize}
Plotting all three groups together reframes group 2 as HB and NB steps, with a transition on the hard apex (similar to behaviors seen in Sco X-1).  The HB and NB data in groups 1 and 3 would then also be steps occurring at a wide variety of values. 

The multi-valued correlation between the optical and Z track position in group 3 became clearer on shorter timescales.  When looking at data covering about an hour, the transition from NB to FB occurred at a distinct step at the soft apex, accompanied by an increased optical scatter in the FB.  The step was much more steep when the system was moving onto the FB rather than off of it.

Timed spectra with 64 s resolution were fit with a two component model that included a blackbody and a cutoff power law, along with a Gaussian to account for the iron K$\alpha$ line.  Fitted parameters tended to behave linearly in each branch, with changes in slope occurring at the apexes.  Each parameter had a unique behavior moving through the Z track, with the exception of the blackbody temperature and the cutoff energy.  In the plots of optical intensity and the fitted parameters, the plots tended to lack consistency between groups.  The powerlaw luminosity was an outlier however, in that in each group, it tended to slide down to lower values as the optical increased.  For all of these reasons, the Z track location remains a better predictor of physical parameters than the optical flux.

\section*{Acknowledgements}

A. B. I. acknowledges support from a Louisiana Board of Regents Fellowship and a Graduate Student Research Assistance award from the Louisiana Space Grant Consortium. This work was also supported by NASA/Louisiana Board of Regents grant NNX07AT62A/LEQSF(2007-10) Phase3-02. This research has made use of data and software provided by the High Energy Astrophysics Science Archive Research Center (HEASARC), which is a service of the Astrophysics Science Division at NASA's Goddard Space Flight Center, as well as NASA’s Astrophysics Data System. We would like to thank the Rossi X-ray Timing Explorer team for their support and especially the schedulers for facilitating simultaneous observations.  This paper includes data taken at The McDonald Observatory of The University of Texas
at Austin, and we thank Amanda Bayless for her assistance in collecting that data.

\section*{Data Availability}

The X-ray data used in this study can be found on HEASARC, using the observation ID 92039.  The optical data are available on Zenodo (DOI: 10.5281/zenodo.11238939).



\bibliographystyle{mnras}
\bibliography{biblio} 

\newcommand{\noop}[1]{}
\begin{thebibliography}{}
\makeatletter
\relax
\def\mn@urlcharsother{\let\do\@makeother \do\$\do\&\do\#\do\^\do\_\do\%\do\~}
\def\mn@doi{\begingroup\mn@urlcharsother \@ifnextchar [ {\mn@doi@}
  {\mn@doi@[]}}
\def\mn@doi@[#1]#2{\def\@tempa{#1}\ifx\@tempa\@empty \href
  {http://dx.doi.org/#2} {doi:#2}\else \href {http://dx.doi.org/#2} {#1}\fi
  \endgroup}
\def\mn@eprint#1#2{\mn@eprint@#1:#2::\@nil}
\def\mn@eprint@arXiv#1{\href {http://arxiv.org/abs/#1} {{\tt arXiv:#1}}}
\def\mn@eprint@dblp#1{\href {http://dblp.uni-trier.de/rec/bibtex/#1.xml}
  {dblp:#1}}
\def\mn@eprint@#1:#2:#3:#4\@nil{\def\@tempa {#1}\def\@tempb {#2}\def\@tempc
  {#3}\ifx \@tempc \@empty \let \@tempc \@tempb \let \@tempb \@tempa \fi \ifx
  \@tempb \@empty \def\@tempb {arXiv}\fi \@ifundefined
  {mn@eprint@\@tempb}{\@tempb:\@tempc}{\expandafter \expandafter \csname
  mn@eprint@\@tempb\endcsname \expandafter{\@tempc}}}

\bibitem[\protect\citeauthoryear{{Ba{\l}uci{\'n}ska-Church}, {Gibiec},
  {Jackson}  \& {Church}}{{Ba{\l}uci{\'n}ska-Church}
  et~al.}{2010}]{2010A&A...512A...9B}
{Ba{\l}uci{\'n}ska-Church} M.,  {Gibiec} A.,  {Jackson} N.~K.,   {Church}
  M.~J.,  2010, \mn@doi [\aap] {10.1051/0004-6361/200913199}, \href
  {https://ui.adsabs.harvard.edu/abs/2010A&A...512A...9B} {512, A9}

\bibitem[\protect\citeauthoryear{{Ba{\l}uci{\'n}ska-Church}, {Church}  \&
  {Gibiec}}{{Ba{\l}uci{\'n}ska-Church} et~al.}{2012}]{2012MmSAI..83..178B}
{Ba{\l}uci{\'n}ska-Church} M.,  {Church} M.~J.,   {Gibiec} A.,  2012, \mn@doi
  [\memsai] {10.48550/arXiv.1110.1266}, \href
  {https://ui.adsabs.harvard.edu/abs/2012MmSAI..83..178B} {83, 178}

\bibitem[\protect\citeauthoryear{{Barret}, {Olive}, {Boirin}, {Done}, {Skinner}
   \& {Grindlay}}{{Barret} et~al.}{2000}]{2000ApJ...533..329B}
{Barret} D.,  {Olive} J.~F.,  {Boirin} L.,  {Done} C.,  {Skinner} G.~K.,
  {Grindlay} J.~E.,  2000, \mn@doi [\apj] {10.1086/308651}, \href
  {https://ui.adsabs.harvard.edu/abs/2000ApJ...533..329B} {533, 329}

\bibitem[\protect\citeauthoryear{{Byram}, {Chubb}  \& {Friedman}}{{Byram}
  et~al.}{1966}]{1966Sci...152...66B}
{Byram} E.~T.,  {Chubb} T.~A.,   {Friedman} H.,  1966, \mn@doi [Science]
  {10.1126/science.152.3718.66}, \href
  {https://ui.adsabs.harvard.edu/abs/1966Sci...152...66B} {152, 66}

\bibitem[\protect\citeauthoryear{{Church} \&
  {Ba{\l}uci{\'n}ska-Church}}{{Church} \&
  {Ba{\l}uci{\'n}ska-Church}}{1995}]{1995A&A...300..441C}
{Church} M.~J.,  {Ba{\l}uci{\'n}ska-Church} M.,  1995, \aap, \href
  {https://ui.adsabs.harvard.edu/abs/1995A&A...300..441C} {300, 441}

\bibitem[\protect\citeauthoryear{{Church} \&
  {Ba{\l}uci{\'n}ska-Church}}{{Church} \&
  {Ba{\l}uci{\'n}ska-Church}}{2004}]{2004MNRAS.348..955C}
{Church} M.~J.,  {Ba{\l}uci{\'n}ska-Church} M.,  2004, \mn@doi [\mnras]
  {10.1111/j.1365-2966.2004.07162.x}, \href
  {https://ui.adsabs.harvard.edu/abs/2004MNRAS.348..955C} {348, 955}

\bibitem[\protect\citeauthoryear{{Church}, {Halai}  \&
  {Ba{\l}uci{\'n}ska-Church}}{{Church} et~al.}{2006}]{2006A&A...460..233C}
{Church} M.~J.,  {Halai} G.~S.,   {Ba{\l}uci{\'n}ska-Church} M.,  2006, \mn@doi
  [\aap] {10.1051/0004-6361:20065035}, \href
  {https://ui.adsabs.harvard.edu/abs/2006A&A...460..233C} {460, 233}

\bibitem[\protect\citeauthoryear{{Cominsky}, {London}  \& {Klein}}{{Cominsky}
  et~al.}{1987}]{1987ApJ...315..162C}
{Cominsky} L.~R.,  {London} R.~A.,   {Klein} R.~I.,  1987, \mn@doi [\apj]
  {10.1086/165122}, \href
  {https://ui.adsabs.harvard.edu/abs/1987ApJ...315..162C} {315, 162}

\bibitem[\protect\citeauthoryear{{Cowley}, {Crampton}  \& {Hutchings}}{{Cowley}
  et~al.}{1979}]{1979ApJ...231..539C}
{Cowley} A.~P.,  {Crampton} D.,   {Hutchings} J.~B.,  1979, \mn@doi [\apj]
  {10.1086/157216}, \href
  {https://ui.adsabs.harvard.edu/abs/1979ApJ...231..539C} {231, 539}

\bibitem[\protect\citeauthoryear{{Devasia}, {Raman}  \& {Paul}}{{Devasia}
  et~al.}{2021}]{2021NewA...8301479D}
{Devasia} J.,  {Raman} G.,   {Paul} B.,  2021, \mn@doi [\na]
  {10.1016/j.newast.2020.101479}, \href
  {https://ui.adsabs.harvard.edu/abs/2021NewA...8301479D} {83, 101479}

\bibitem[\protect\citeauthoryear{{Dieters} \& {van der Klis}}{{Dieters} \& {van
  der Klis}}{2000}]{2000MNRAS.311..201D}
{Dieters} S.~W.,  {van der Klis} M.,  2000, \mn@doi [\mnras]
  {10.1046/j.1365-8711.2000.03050.x10.48550/arXiv.astro-ph/9909472}, \href
  {https://ui.adsabs.harvard.edu/abs/2000MNRAS.311..201D} {311, 201}

\bibitem[\protect\citeauthoryear{{Ding} \& {Huang}}{{Ding} \&
  {Huang}}{2015}]{2015JApA...36..335D}
{Ding} G.~Q.,  {Huang} C.~P.,  2015, \mn@doi [Journal of Astrophysics and
  Astronomy] {10.1007/s12036-015-9340-2}, \href
  {https://ui.adsabs.harvard.edu/abs/2015JApA...36..335D} {36, 335}

\bibitem[\protect\citeauthoryear{{Ding}, {Deller}  \& {Miller-Jones}}{{Ding}
  et~al.}{2021}]{2021PASA...38...48D}
{Ding} H.,  {Deller} A.~T.,   {Miller-Jones} J. C.~A.,  2021, \mn@doi [\pasa]
  {10.1017/pasa.2021.37}, \href
  {https://ui.adsabs.harvard.edu/abs/2021PASA...38...48D} {38, e048}

\bibitem[\protect\citeauthoryear{{Edelson} \& {Krolik}}{{Edelson} \&
  {Krolik}}{1988}]{1988ApJ...333..646E}
{Edelson} R.~A.,  {Krolik} J.~H.,  1988, \mn@doi [\apj] {10.1086/166773}, \href
  {https://ui.adsabs.harvard.edu/abs/1988ApJ...333..646E} {333, 646}

\bibitem[\protect\citeauthoryear{Frenet}{Frenet}{1852}]{Frenet1852}
Frenet F.,  1852, Journal de Mathématiques Pures et Appliquées, pp 437--447

\bibitem[\protect\citeauthoryear{{Fridriksson}, {Homan}  \&
  {Remillard}}{{Fridriksson} et~al.}{2015}]{2015ApJ...809...52F}
{Fridriksson} J.~K.,  {Homan} J.,   {Remillard} R.~A.,  2015, \mn@doi [\apj]
  {10.1088/0004-637X/809/1/52}, \href
  {https://ui.adsabs.harvard.edu/abs/2015ApJ...809...52F} {809, 52}

\bibitem[\protect\citeauthoryear{{Gaskell} \& {Peterson}}{{Gaskell} \&
  {Peterson}}{1987}]{1987ApJS...65....1G}
{Gaskell} C.~M.,  {Peterson} B.~M.,  1987, \mn@doi [\apjs] {10.1086/191216},
  \href {https://ui.adsabs.harvard.edu/abs/1987ApJS...65....1G} {65, 1}

\bibitem[\protect\citeauthoryear{{Hasinger} \& {van der Klis}}{{Hasinger} \&
  {van der Klis}}{1989}]{1989A&A...225...79H}
{Hasinger} G.,  {van der Klis} M.,  1989, \aap, \href
  {https://ui.adsabs.harvard.edu/abs/1989A&A...225...79H} {225, 79}

\bibitem[\protect\citeauthoryear{{Hasinger}, {van der Klis}, {Ebisawa},
  {Dotani}  \& {Mitsuda}}{{Hasinger} et~al.}{1990}]{1990A&A...235..131H}
{Hasinger} G.,  {van der Klis} M.,  {Ebisawa} K.,  {Dotani} T.,   {Mitsuda} K.,
   1990, \aap, \href {https://ui.adsabs.harvard.edu/abs/1990A&A...235..131H}
  {235, 131}

\bibitem[\protect\citeauthoryear{{Hertz}, {Vaughan}, {Wood}, {Norris},
  {Mitsuda}, {Michelson}  \& {Dotani}}{{Hertz}
  et~al.}{1992}]{1992ApJ...396..201H}
{Hertz} P.,  {Vaughan} B.,  {Wood} K.~S.,  {Norris} J.~P.,  {Mitsuda} K.,
  {Michelson} P.~F.,   {Dotani} T.,  1992, \mn@doi [\apj] {10.1086/171710},
  \href {https://ui.adsabs.harvard.edu/abs/1992ApJ...396..201H} {396, 201}

\bibitem[\protect\citeauthoryear{{Homan} et~al.,}{{Homan}
  et~al.}{2007a}]{2007ApJ...656..420H}
{Homan} J.,  et~al., 2007a, \mn@doi [\apj] {10.1086/510447}, \href
  {https://ui.adsabs.harvard.edu/abs/2007ApJ...656..420H} {656, 420}

\bibitem[\protect\citeauthoryear{{Homan}, {Belloni}, {Wijnands}, {van der
  Klis}, {Swank}, {Smith}, {Pereira}  \& {Markwardt}}{{Homan}
  et~al.}{2007b}]{2007ATel.1144....1H}
{Homan} J.,  {Belloni} T.,  {Wijnands} R.,  {van der Klis} M.,  {Swank} J.,
  {Smith} E.,  {Pereira} D.,   {Markwardt} C.,  2007b, The Astronomer's
  Telegram, \href {https://ui.adsabs.harvard.edu/abs/2007ATel.1144....1H}
  {1144, 1}

\bibitem[\protect\citeauthoryear{{Horne}}{{Horne}}{2003}]{2003SPIE.4854..262H}
{Horne} K.,  2003, in {Blades} J.~C.,  {Siegmund} O. H.~W.,  eds,  Society of
  Photo-Optical Instrumentation Engineers (SPIE) Conference Series Vol. 4854,
  Future EUV/UV and Visible Space Astrophysics Missions and Instrumentation..
  pp 262--273 (\mn@eprint {arXiv} {astro-ph/0301250}),
  \mn@doi{10.1117/12.460261}

\bibitem[\protect\citeauthoryear{{Hynes}, {Schaefer}, {Baum}, {Hsu}, {Cherry}
  \& {Scaringi}}{{Hynes} et~al.}{2016}]{2016MNRAS.459.3596H}
{Hynes} R.~I.,  {Schaefer} B.~E.,  {Baum} Z.~A.,  {Hsu} C.-C.,  {Cherry} M.~L.,
    {Scaringi} S.,  2016, \mn@doi [\mnras] {10.1093/mnras/stw854}, \href
  {https://ui.adsabs.harvard.edu/abs/2016MNRAS.459.3596H} {459, 3596}

\bibitem[\protect\citeauthoryear{Igl}{Igl}{2023}]{igl_2023}
Igl A.,  2023, PhD thesis, Louisiana State University

\bibitem[\protect\citeauthoryear{{Igl}, {Hynes}, {Britt}, {O'Brien}  \&
  {Mikles}}{{Igl} et~al.}{2023}]{igl_scox1}
{Igl} A.~B.,  {Hynes} R.~I.,  {Britt} C.~T.,  {O'Brien} K.~S.,   {Mikles}
  V.~J.,  2023, \mn@doi [arXiv e-prints] {10.48550/arXiv.2308.12934}, \href
  {https://ui.adsabs.harvard.edu/abs/2023arXiv230812934I} {p. arXiv:2308.12934}

\bibitem[\protect\citeauthoryear{{Kuulkers}, {van der Klis}, {Oosterbroek},
  {Asai}, {Dotani}, {van Paradijs}  \& {Lewin}}{{Kuulkers}
  et~al.}{1994}]{1994A&A...289..795K}
{Kuulkers} E.,  {van der Klis} M.,  {Oosterbroek} T.,  {Asai} K.,  {Dotani} T.,
   {van Paradijs} J.,   {Lewin} W.~H.~G.,  1994, \aap, \href
  {https://ui.adsabs.harvard.edu/abs/1994A&A...289..795K} {289, 795}

\bibitem[\protect\citeauthoryear{{Kuulkers}, {van der Klis}  \&
  {Vaughan}}{{Kuulkers} et~al.}{1996}]{1996A&A...311..197K}
{Kuulkers} E.,  {van der Klis} M.,   {Vaughan} B.~A.,  1996, \aap, \href
  {https://ui.adsabs.harvard.edu/abs/1996A&A...311..197K} {311, 197}

\bibitem[\protect\citeauthoryear{{Lin}, {Remillard}  \& {Homan}}{{Lin}
  et~al.}{2009}]{2009ApJ...696.1257L}
{Lin} D.,  {Remillard} R.~A.,   {Homan} J.,  2009, \mn@doi [\apj]
  {10.1088/0004-637X/696/2/1257}, \href
  {https://ui.adsabs.harvard.edu/abs/2009ApJ...696.1257L} {696, 1257}

\bibitem[\protect\citeauthoryear{{Ludlam} et~al.,}{{Ludlam}
  et~al.}{2022}]{2022ApJ...927..112L}
{Ludlam} R.~M.,  et~al., 2022, \mn@doi [\apj] {10.3847/1538-4357/ac5028}, \href
  {https://ui.adsabs.harvard.edu/abs/2022ApJ...927..112L} {927, 112}

\bibitem[\protect\citeauthoryear{{Mitsuda}, {Inoue}, {Nakamura}  \&
  {Tanaka}}{{Mitsuda} et~al.}{1989}]{1989PASJ...41...97M}
{Mitsuda} K.,  {Inoue} H.,  {Nakamura} N.,   {Tanaka} Y.,  1989, \pasj, \href
  {https://ui.adsabs.harvard.edu/abs/1989PASJ...41...97M} {41, 97}

\bibitem[\protect\citeauthoryear{{Mondal}, {Dewangan}, {Pahari}  \&
  {Raychaudhuri}}{{Mondal} et~al.}{2018}]{2018MNRAS.474.2064M}
{Mondal} A.~S.,  {Dewangan} G.~C.,  {Pahari} M.,   {Raychaudhuri} B.,  2018,
  \mn@doi [\mnras] {10.1093/mnras/stx2931}, \href
  {https://ui.adsabs.harvard.edu/abs/2018MNRAS.474.2064M} {474, 2064}

\bibitem[\protect\citeauthoryear{{Mukadam} \& {Nather}}{{Mukadam} \&
  {Nather}}{2005}]{2005JApA...26..321M}
{Mukadam} A.~S.,  {Nather} R.~E.,  2005, \mn@doi [Journal of Astrophysics and
  Astronomy] {10.1007/BF02702340}, \href
  {https://ui.adsabs.harvard.edu/abs/2005JApA...26..321M} {26, 321}

\bibitem[\protect\citeauthoryear{O'Brien}{O'Brien}{2000}]{obrien_2000}
O'Brien K.,  2000, PhD thesis, University of St. Andrews

\bibitem[\protect\citeauthoryear{{O'Brien}, {Horne}, {Hynes}, {Chen}, {Haswell}
   \& {Still}}{{O'Brien} et~al.}{2002}]{2002MNRAS.334..426O}
{O'Brien} K.,  {Horne} K.,  {Hynes} R.~I.,  {Chen} W.,  {Haswell} C.~A.,
  {Still} M.~D.,  2002, \mn@doi [\mnras] {10.1046/j.1365-8711.2002.05530.x},
  \href {https://ui.adsabs.harvard.edu/abs/2002MNRAS.334..426O} {334, 426}

\bibitem[\protect\citeauthoryear{{O'Brien}, {Horne}, {Gomer}, {Oke}  \& {van
  der Klis}}{{O'Brien} et~al.}{2004}]{2004MNRAS.350..587O}
{O'Brien} K.,  {Horne} K.,  {Gomer} R.~H.,  {Oke} J.~B.,   {van der Klis} M.,
  2004, \mn@doi [\mnras] {10.1111/j.1365-2966.2004.07667.x}, \href
  {https://ui.adsabs.harvard.edu/abs/2004MNRAS.350..587O} {350, 587}

\bibitem[\protect\citeauthoryear{{Orosz} \& {Kuulkers}}{{Orosz} \&
  {Kuulkers}}{1999}]{1999MNRAS.305..132O}
{Orosz} J.~A.,  {Kuulkers} E.,  1999, \mn@doi [\mnras]
  {10.1046/j.1365-8711.1999.t01-1-02420.x}, \href
  {https://ui.adsabs.harvard.edu/abs/1999MNRAS.305..132O} {305, 132}

\bibitem[\protect\citeauthoryear{{Pringle}}{{Pringle}}{1997}]{1997MNRAS.292..136P}
{Pringle} J.~E.,  1997, \mn@doi [\mnras] {10.1093/mnras/292.1.136}, \href
  {https://ui.adsabs.harvard.edu/abs/1997MNRAS.292..136P} {292, 136}

\bibitem[\protect\citeauthoryear{{Psaltis}, {Lamb}  \& {Miller}}{{Psaltis}
  et~al.}{1995}]{1995ApJ...454L.137P}
{Psaltis} D.,  {Lamb} F.~K.,   {Miller} G.~S.,  1995, \mn@doi [\apjl]
  {10.1086/309780}, \href
  {https://ui.adsabs.harvard.edu/abs/1995ApJ...454L.137P} {454, L137}

\bibitem[\protect\citeauthoryear{{Remillard}, {Lin}, {ASM Team at MIT}  \&
  {NASA/GSFC}}{{Remillard} et~al.}{2006}]{2006ATel..696....1R}
{Remillard} R.~A.,  {Lin} D.,  {ASM Team at MIT}  {NASA/GSFC} 2006, The
  Astronomer's Telegram, \href
  {https://ui.adsabs.harvard.edu/abs/2006ATel..696....1R} {696, 1}

\bibitem[\protect\citeauthoryear{{Rhodes}, {Fender}, {Motta}, {van den
  Eijnden}, {Williams}, {Bright}  \& {Sivakoff}}{{Rhodes}
  et~al.}{2022}]{2022MNRAS.513.2708R}
{Rhodes} L.,  {Fender} R.~P.,  {Motta} S.,  {van den Eijnden} J.,  {Williams}
  D.~R.~A.,  {Bright} J.,   {Sivakoff} G.~R.,  2022, \mn@doi [\mnras]
  {10.1093/mnras/stac954}, \href
  {https://ui.adsabs.harvard.edu/abs/2022MNRAS.513.2708R} {513, 2708}

\bibitem[\protect\citeauthoryear{{Scaringi}, {Maccarone}, {Hynes},
  {K{\"o}rding}, {Ponti}, {Knigge}, {Britt}  \& {van Winckel}}{{Scaringi}
  et~al.}{2015}]{2015MNRAS.451.3857S}
{Scaringi} S.,  {Maccarone} T.~J.,  {Hynes} R.~I.,  {K{\"o}rding} E.,  {Ponti}
  G.,  {Knigge} C.,  {Britt} C.~T.,   {van Winckel} H.,  2015, \mn@doi [\mnras]
  {10.1093/mnras/stv1216}, \href
  {https://ui.adsabs.harvard.edu/abs/2015MNRAS.451.3857S} {451, 3857}

\bibitem[\protect\citeauthoryear{{Schulz}, {Huenemoerder}, {Ji}, {Nowak}, {Yao}
   \& {Canizares}}{{Schulz} et~al.}{2009}]{2009ApJ...692L..80S}
{Schulz} N.~S.,  {Huenemoerder} D.~P.,  {Ji} L.,  {Nowak} M.,  {Yao} Y.,
  {Canizares} C.~R.,  2009, \mn@doi [\apjl] {10.1088/0004-637X/692/2/L80},
  \href {https://ui.adsabs.harvard.edu/abs/2009ApJ...692L..80S} {692, L80}

\bibitem[\protect\citeauthoryear{Serret}{Serret}{1851}]{Serret1851}
Serret J.-A.,  1851, Journal de Mathématiques Pures et Appliquées, pp
  193--207

\bibitem[\protect\citeauthoryear{{Smale} et~al.,}{{Smale}
  et~al.}{1993}]{1993ApJ...410..796S}
{Smale} A.~P.,  et~al., 1993, \mn@doi [\apj] {10.1086/172796}, \href
  {https://ui.adsabs.harvard.edu/abs/1993ApJ...410..796S} {410, 796}

\bibitem[\protect\citeauthoryear{{Titarchuk}, {Kuznetsov}  \&
  {Shaposhnikov}}{{Titarchuk} et~al.}{2007}]{2007ApJ...667..404T}
{Titarchuk} L.,  {Kuznetsov} S.,   {Shaposhnikov} N.,  2007, \mn@doi [\apj]
  {10.1086/520759}, \href
  {https://ui.adsabs.harvard.edu/abs/2007ApJ...667..404T} {667, 404}

\bibitem[\protect\citeauthoryear{{Vrtilek}, {Kahn}, {Grindlay}, {Helfand}  \&
  {Seward}}{{Vrtilek} et~al.}{1986}]{1986ApJ...307..698V}
{Vrtilek} S.~D.,  {Kahn} S.~M.,  {Grindlay} J.~E.,  {Helfand} D.~J.,   {Seward}
  F.~D.,  1986, \mn@doi [\apj] {10.1086/164455}, \href
  {https://ui.adsabs.harvard.edu/abs/1986ApJ...307..698V} {307, 698}

\bibitem[\protect\citeauthoryear{{Vrtilek}, {Swank}, {Kelley}  \&
  {Kahn}}{{Vrtilek} et~al.}{1988}]{1988ApJ...329..276V}
{Vrtilek} S.~D.,  {Swank} J.~H.,  {Kelley} R.~L.,   {Kahn} S.~M.,  1988,
  \mn@doi [\apj] {10.1086/166376}, \href
  {https://ui.adsabs.harvard.edu/abs/1988ApJ...329..276V} {329, 276}

\bibitem[\protect\citeauthoryear{{Vrtilek}, {Raymond}, {Boroson}, {McCray},
  {Smale}, {Kallman}  \& {Nagase}}{{Vrtilek}
  et~al.}{2003}]{2003PASP..115.1124V}
{Vrtilek} S.~D.,  {Raymond} J.~C.,  {Boroson} B.,  {McCray} R.,  {Smale} A.,
  {Kallman} T.,   {Nagase} F.,  2003, \mn@doi [\pasp] {10.1086/377089}, \href
  {https://ui.adsabs.harvard.edu/abs/2003PASP..115.1124V} {115, 1124}

\bibitem[\protect\citeauthoryear{{White} \& {Swank}}{{White} \&
  {Swank}}{1982}]{1982ApJ...253L..61W}
{White} N.~E.,  {Swank} J.~H.,  1982, \mn@doi [\apjl] {10.1086/183737}, \href
  {https://ui.adsabs.harvard.edu/abs/1982ApJ...253L..61W} {253, L61}

\bibitem[\protect\citeauthoryear{{van Paradijs} \& {McClintock}}{{van Paradijs}
  \& {McClintock}}{1994}]{1994A&A...290..133V}
{van Paradijs} J.,  {McClintock} J.~E.,  1994, \aap, \href
  {https://ui.adsabs.harvard.edu/abs/1994A&A...290..133V} {290, 133}

\makeatother
\end{thebibliography}




\newpage
\appendix

\clearpage
\section{Overlapping Lightcurves by Modified Julian Day}
\label{lc_apx}

As this Cyg X-2 data has not yet been published, the following appendix contains every overlapping X-ray and optical lightcurve, organized by MJD.  The optical data are differential, divided by the median of every differential lightcurve.  All axes have the same scaling for ease of comparison.

\begin{figure}
    \setlength\tabcolsep{0pt}
    \begin{tabular}{l}
        \begin{subfigure}[b]{.475\textwidth}
            \setcounter{subfigure}{0}
            \includegraphics[width=\textwidth]{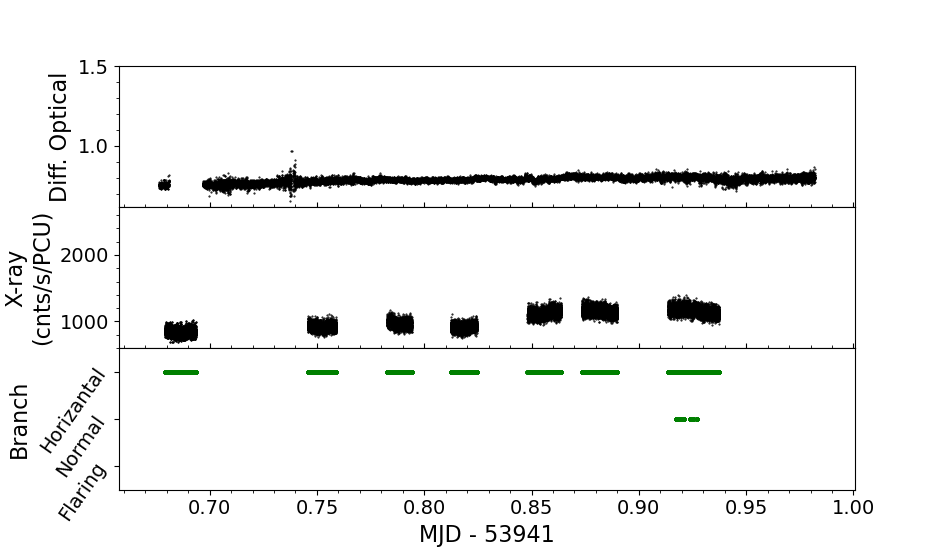}
            \caption[empty]{}
            \label{fig:53941_lc}
        \end{subfigure}\\
        \hfill
        \begin{subfigure}[b]{.475\textwidth}
            \setcounter{subfigure}{1}
            \includegraphics[width=\textwidth]{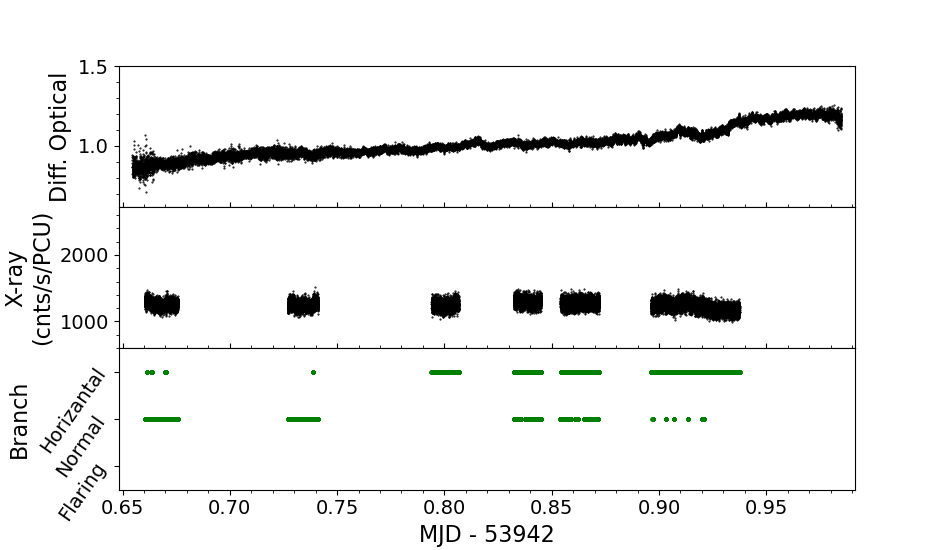}
            \caption[empty]{}
            \label{fig:53942_lc}
        \end{subfigure}\\
        \hfill
        \begin{subfigure}[b]{.475\textwidth}
            \setcounter{subfigure}{2}
            \includegraphics[width=\textwidth]{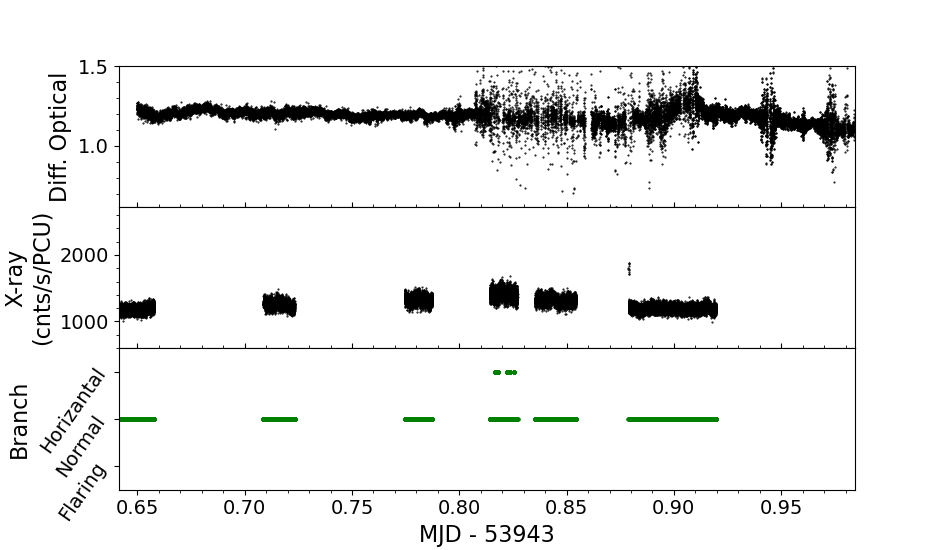}
            \caption[empty]{}
            \label{fig:53943_lc}
        \end{subfigure}\\
        \hfill
        \begin{subfigure}[b]{.475\textwidth}
            \setcounter{subfigure}{3}
            \includegraphics[width=\textwidth]{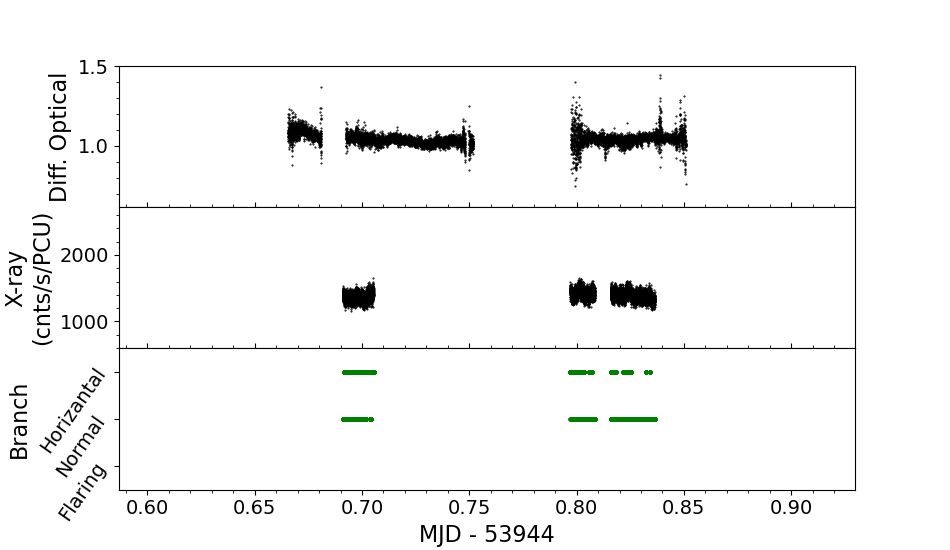}
            \caption[empty]{}
            \label{fig:may25a_lc}
        \end{subfigure}
    \end{tabular}
    \caption{The top plot is the differential optical lightcurve (normalized by the median), the middle is the X-ray intensity, and the bottom is the location on the Z track.}
\end{figure}

\begin{figure*}
    \ContinuedFloat
    \setlength\tabcolsep{7.5pt}
    \begin{tabular}{cc}
        \begin{subfigure}[b]{.475\textwidth}
            \setcounter{subfigure}{4}
            \includegraphics[width=\textwidth]{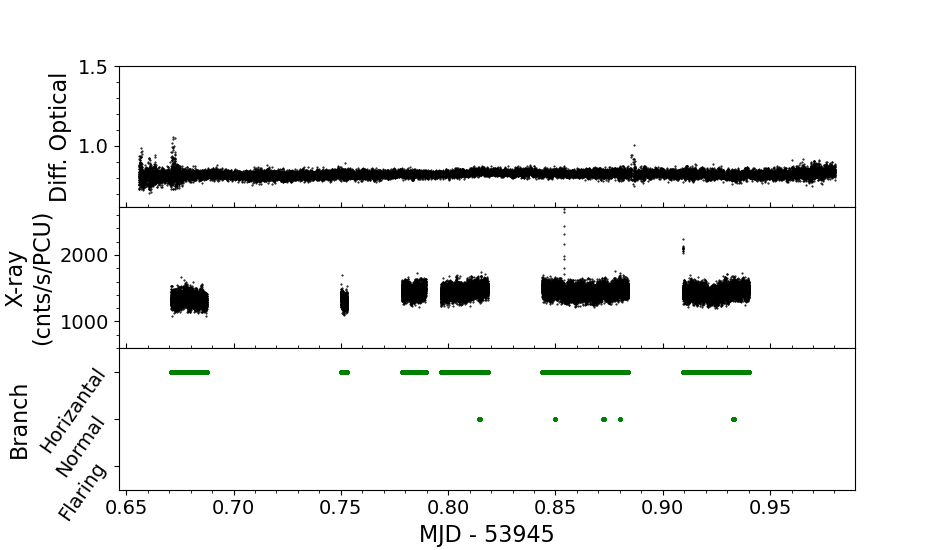}
            \caption[empty]{}
            \label{fig:53945_lc}
        \end{subfigure}&
        \hfill
        \begin{subfigure}[b]{.475\textwidth}
            \setcounter{subfigure}{8}
            \includegraphics[width=\textwidth]{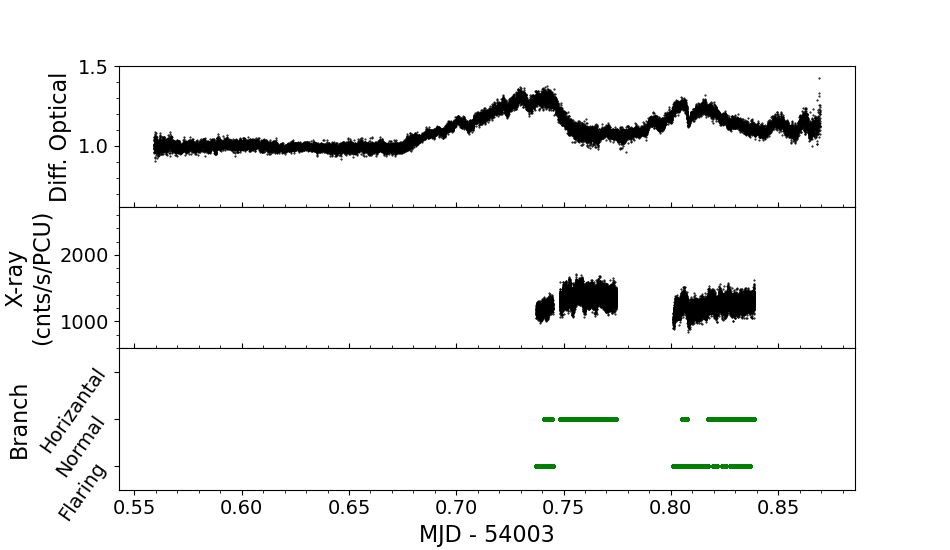}
            \caption[empty]{}
            \label{fig:54003_lc}
        \end{subfigure}\\
        \hfill
        \begin{subfigure}[b]{.475\textwidth}
            \setcounter{subfigure}{5}
            \includegraphics[width=\textwidth]{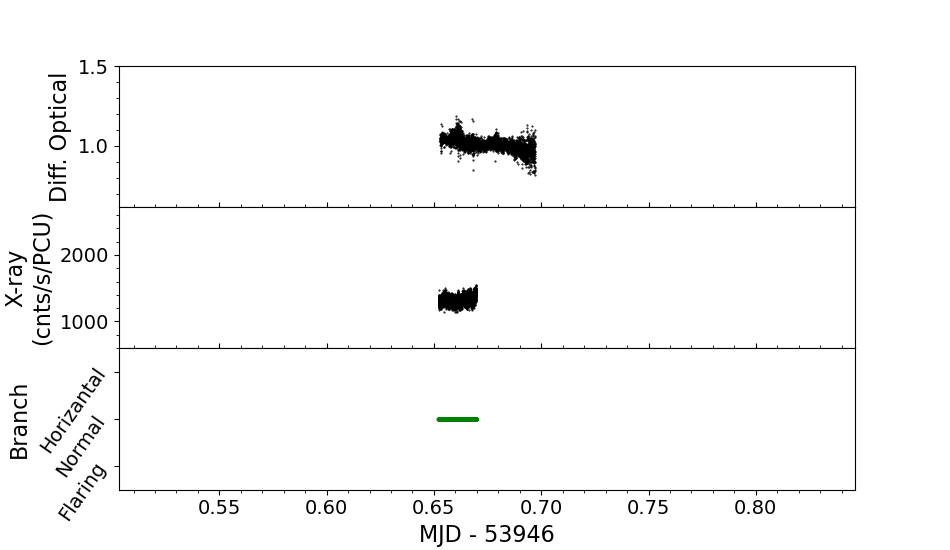}
            \caption[empty]{}
            \label{fig:53946_lc}
        \end{subfigure}&
        \hfill
        \begin{subfigure}[b]{.475\textwidth}
            \setcounter{subfigure}{9}
            \includegraphics[width=\textwidth]{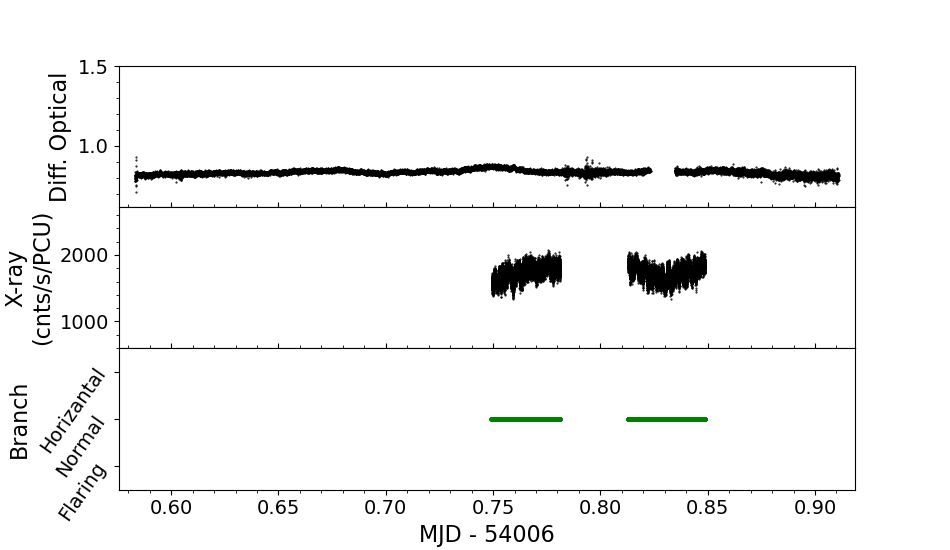}
            \caption[empty]{}
            \label{fig:54006_lc}
        \end{subfigure}\\
        \hfill
        \begin{subfigure}[b]{.475\textwidth}
            \setcounter{subfigure}{6}
            \includegraphics[width=\textwidth]{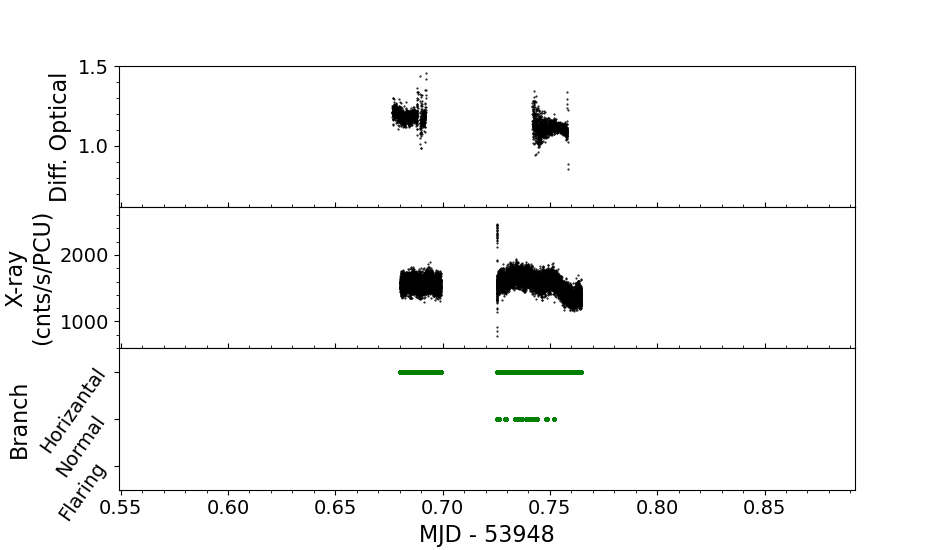}
            \caption[empty]{}
            \label{fig:53948_lc}
        \end{subfigure}&
        \hfill
        \begin{subfigure}[b]{.475\textwidth}
            \setcounter{subfigure}{10}
            \includegraphics[width=\textwidth]{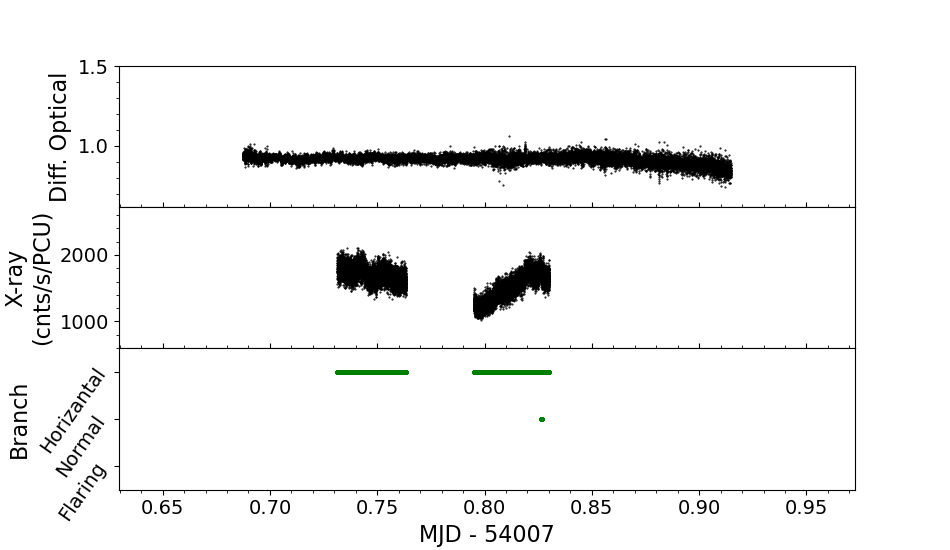}
            \caption[empty]{}
            \label{fig:54007_lc}
        \end{subfigure}\\
        \hfill
        \begin{subfigure}[b]{.475\textwidth}
            \setcounter{subfigure}{7}
            \includegraphics[width=\textwidth]{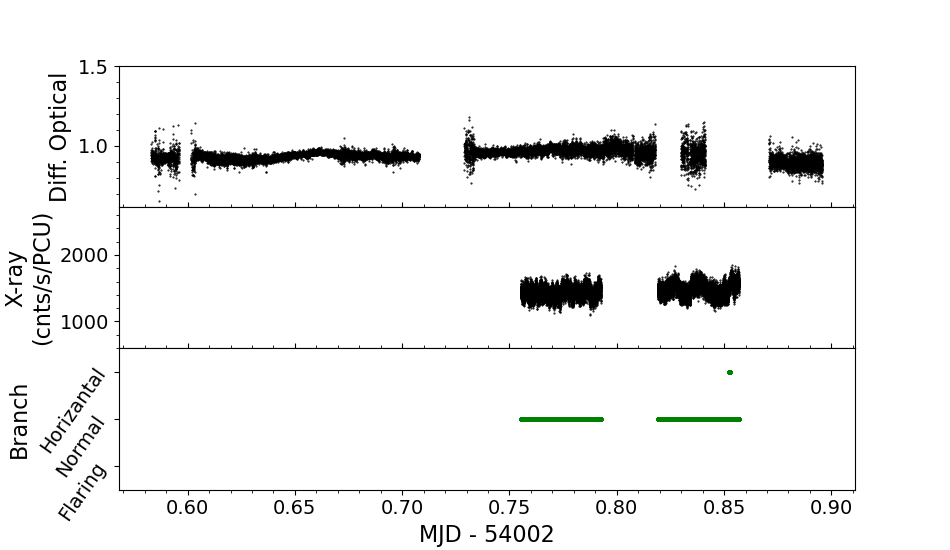}
            \caption[empty]{}
            \label{fig:54002_lc}
        \end{subfigure}&
        \begin{subfigure}[b]{.475\textwidth}
            \setcounter{subfigure}{11}
            \includegraphics[width=\textwidth]{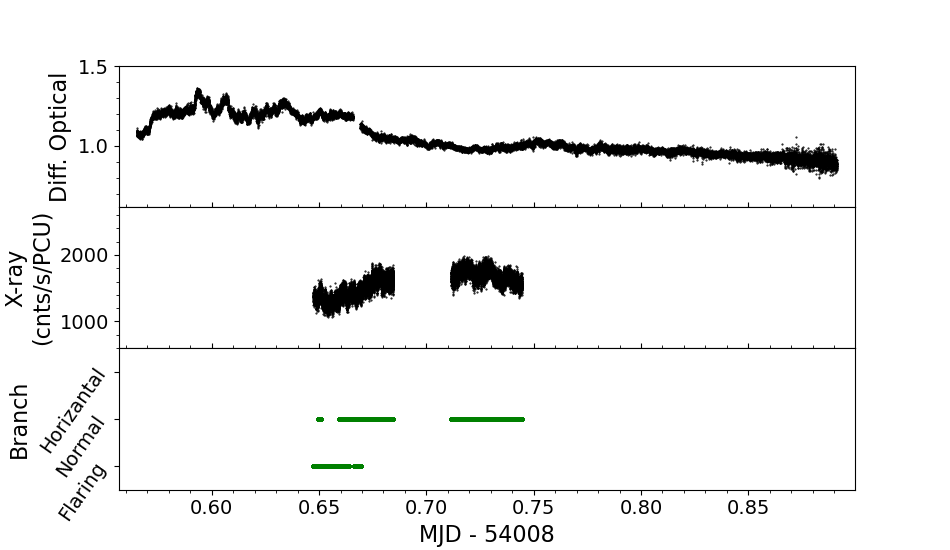}
            \caption[empty]{}
            \label{fig:54008_lc}
        \end{subfigure}
        \hfill
    \end{tabular}
    \caption{(cont.) The top plot is the differential optical lightcurve (normalized by the median), the middle is the X-ray intensity, and the bottom is the location on the Z track.}
\end{figure*}

\begin{figure}
    \ContinuedFloat
    \setlength\tabcolsep{7.5pt}
    \begin{tabular}{l}
        \begin{subfigure}[b]{.475\textwidth}
            \setcounter{subfigure}{12}
            \includegraphics[width=\textwidth]{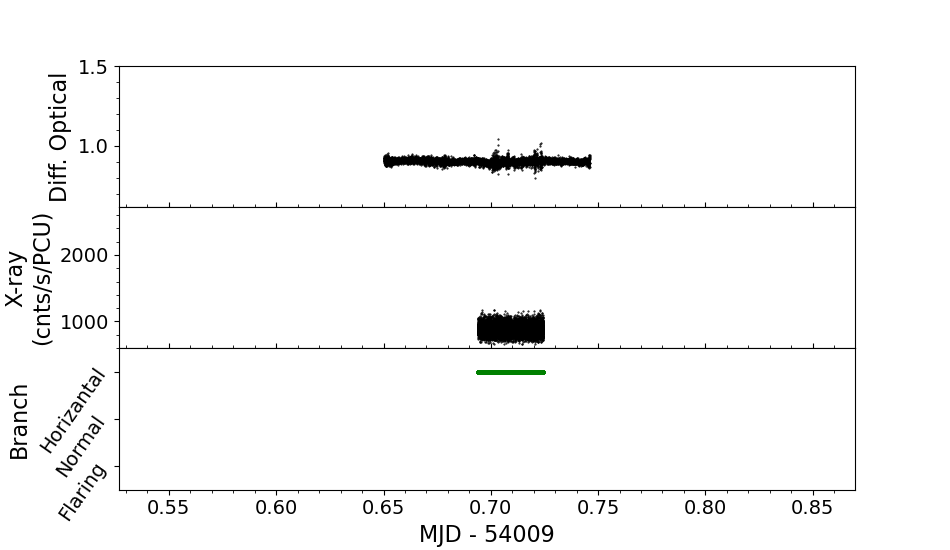}
            \caption[empty]{}
            \label{fig:54009_lc}
        \end{subfigure}\\
        \hfill
        \begin{subfigure}[b]{.475\textwidth}
            \setcounter{subfigure}{13}
            \includegraphics[width=\textwidth]{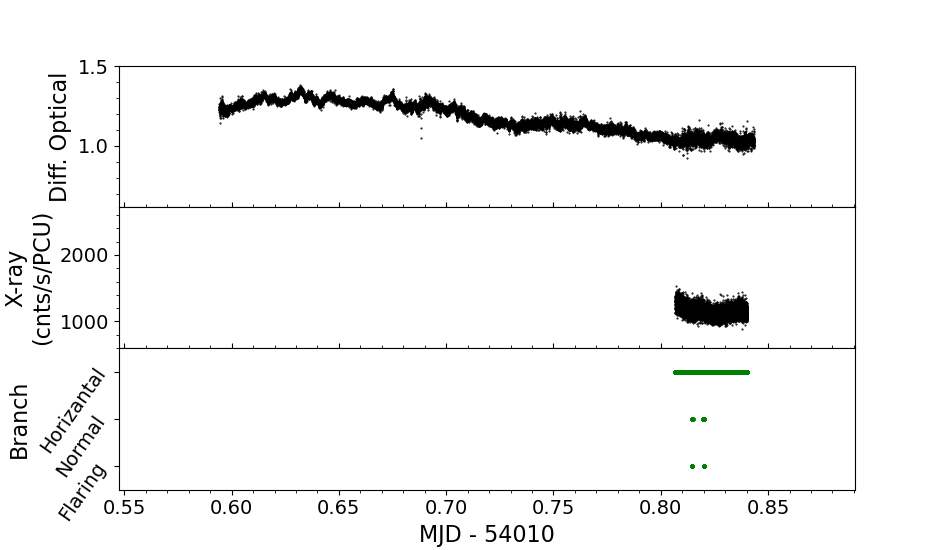}
            \caption[empty]{}
            \label{fig:54010_lc}
        \end{subfigure}\\
    \end{tabular}
    \caption{(cont.) The top plot is the differential optical lightcurve (normalized by the median), the middle is the X-ray intensity, and the bottom is the location on the Z track.}
\end{figure}

\bsp	
\label{lastpage}
\end{document}